\documentclass[twoside,fleqn]{article}
\usepackage{amsfonts,gc97lanl}

\def\RR{{\Bbb R}}

\def\CC{{\Bbb C}}

\def\ZZ{{\Bbb Z}}

\def\stackunder#1#2{\mathrel{\mathop{#2}\limits_{#1}}}
\catcode`\@=11
\def\Let@{\relax\iffalse{\fi\let\\=\cr\iffalse}\fi}
\def\vspace@{\def\vspace##1{\crcr\noalign{\vskip##1\relax}}}
\def\multilimits@{\bgroup\vspace@\Let@
 \baselineskip\fontdimen10 \scriptfont\tw@
 \advance\baselineskip\fontdimen12 \scriptfont\tw@
 \lineskip\thr@@\fontdimen8 \scriptfont\thr@@
 \lineskiplimit\lineskip
 \vbox\bgroup\ialign\bgroup\hfil$\m@th\scriptstyle{##}$\hfil\crcr}
\def\Sb{_\multilimits@}
\def\endSb{\crcr\egroup\egroup\egroup}
\def\Sp{^\multilimits@}

\begin{document}

\jheads{U.G\"unther, S.Kriskiv, A.Zhuk}
       {On stable compactification with Casimir-like potential}

\twocolumn[
\LANL{gr-qc/9801013}
\Title{ON STABLE COMPACTIFICATION WITH CASIMIR-LIKE POTENTIAL}

\Author{U.G\"unther\foom {1}, S.Kriskiv, A.Zhuk\foom {2}}
     {Department of Theoretical Physics, University of Odessa,
     2 Petra Velikogo Str., 270100 Odessa, Ukraine}

\Abstract{
Multidimensional cosmological models with a higher
dimensional space-time manifold $M = \RR \times M_0\times\prod\nolimits_{i=1}
^nM_i$  $( n>1)$  are investigated under dimensional reduction
to $D_0$-dimensional effective models. In the Einstein conformal
frame, the effective potential for the internal scale factors is
obtained. The stable compactification of the internal spaces is
achieved due to the Casimir effect. In the case of more than one internal
space a Casimir-like ansatz for the energy density of the massless
scalar field fluctuations is proposed. Stable configurations with
respect to the internal scale factor excitations are found in the
cases of one and two internal spaces.
}
]  %%%%%%%%%%%%%%%%%%% end of temporary one-column mode
\email{1}{guenther@pool.hrz.htw-zittau.de}
\email{2}{zhuk@paco.odessa.ua}

\hspace*{0.950cm} PACS number(s): 04.50.+h, 98.80.Hw 
%%%%%%%%%%%%%%%%%%%%%%%%%%%%%%%%%%%%%%%%%%%%%%%%%%%%%%%%%%%%%%%%%

\section{Introduction}

\setcounter{equation}{0}
\renewcommand{\theequation}{\thesection\arabic{equation}}

Our observable universe at the present time at large scales is well
described by the Friedmann model with 4-dimensional
Friedmann-Robertson-Walker metric. However, it is possible that space-time
at short (Planck) distances might have a dimensionality of more than four
and possess a rather complex topology \cite{W}. Accounting this possibility
it is natural to generalize the Friedmann model to a multidimensional
cosmological model (MCM) with the topology \cite{GRT}, \cite{IMZ} 
\begin{equation}
\label{1.1}M=\RR \times M_0\times M_1\times \ldots \times M_n , 
\end{equation}
where $M_i\ (i=0,\ldots ,n)$ are spaces of constant curvature (more
generally, Einstein spaces). $M_0$ usually denotes a $d_0$-dimensional
external space and $M_i\ (i=1,\ldots ,n)$ are $d_i$-dimensional internal
spaces. These internal spaces have to be compact, what can be achieved by
appropriate periodicity conditions for the coordinates \cite{Ell}-\cite{NSh}%
. As a result, internal spaces may have a nontrivial topology, being compact
(i.e. closed and bounded) for any sign of the spatial curvature.

To make the internal dimensions unobservable at the present time the
internal spaces have to be reduced to scales near the Planck length $%
L_{Pl}\sim 10^{-33}$cm, i.e. scale factors of the internal spaces $a_i$
should be of order of $L_{Pl}$. Obviously, such compactifications have to be
stable. It means an effective potential of the model obtained under
dimensional reduction to a 4-dimensional effective theory should have minima
at $a_i\sim L_{Pl}\ (i=1,\ldots ,n)$. A lot of papers were devoted to the
problem of stable compactification of extra dimensions (an extended list of
references may be found in the paper \cite{GZ}). A number of effective
potentials which ensure stability was obtained there. Among them the Casimir
potential is one of the most important. The Casimir effect is connected with
the vacuum polarization of quantized fields due to non-trivial topology of
the background space or the presence of boundaries in the space. As a result
one obtains a nonvanishing energy density of the quantized fields in the
vacuum state. In our case this phenomenon should take place due to internal
spaces compactness.

The Casimir energy density for a massless scalar field in a model with one
finite scale factor has a form \cite{Ford}-\cite{KZ2} 
\begin{equation}
\label{1.2}\rho =C\frac 1{a^D}\ , 
\end{equation}
where $D$ is the total space-time dimension and $C$ is a constant which
depends strongly on the topology of the model. Equation (\ref{1.2}) holds
for models with one factor-space (e.g. $M_0$ with the scale factor $%
a_0\equiv a$) \cite{Ford}, \cite{MMS}, \cite{MT1}, \cite{MT2}, \cite{KZ2},
or in the case of one internal space and an external scale factor that is
much greater than the internal one: $a_0\gg a_1\equiv a$ \cite{DeW}-\cite
{Koik}.

In the case of several internal compact spaces the Casimir effect occurs if
at least one of them is finite. It is natural to consider all internal
spaces on equal footing supposing that their scale factors are freezed near
Planck length. In the papers \cite{AGHK,HKVW} an approximation of the
Casimir energy density was proposed in the form 
\begin{equation}
\label{1.3}\rho =\frac 1{\prod_{i=1}^nV_i}\sum_{i=1}^n\frac{A^{(i)}}{a_i^4} 
\end{equation}
for the universe with the topology $M=\RR \times S^3\times \prod_{i=1}^nS^d$
where $V_i\sim a_i^d$ is the volume of the $d$ - dimensional sphere. It can
be easily seen from this equation that $\rho \rightarrow 0$ if any of $%
a_i\rightarrow \infty $. This means that approximation (\ref{1.3}) is not
applicable in this limit because, first, for any finite $a_i$ there should
exist a nonvanishing Casimir energy density, and, second, it should provide
a correct transition to Eq. (\ref{1.2}), what, obviously, is not the case.

The calculation of the Casimir energy density even in the case of one scale
factor is a complicated problem. For several scale factors this procedure
becomes extremely difficult, especially analytically. In the present paper
we suggest ad hoc a Casimir-like ansatz for the energy density of the
massless scalar field fluctuations, which yields a better approximation than
(\ref{1.3}). In this ansatz all internal scale factors are included on equal
footing, it has the correct Casimir dimension: $\mbox{[cm]}^{-D}$ and gives
a correct transition to the formula (\ref{1.2}). The proposed ansatz gives
an energy density that does not equal to zero if at least one of scale
factors is finite.

As it was mentioned above the problem of stable compactification is one of
the most important in MCM. In our paper we investigate this problem for the
proposed Casimir-like potential in the case of one and two internal scale
factors. It is shown that stable configurations exist for both of these
cases. To our knowledge the investigation of the stable compactification
problem due to Casimir potential for models with more than one non-identical
(with respect to the scale factors) spaces was not performed up to now.

The paper is organized as follows. In Section~2, the general description of
the considered model is given. In Section~3, the effective potential is
obtained under dimensional reduction to a $D_0$-dimensional (usually $D_0=4$%
) effective theory in the Einstein frame. The problem of stable
compactification is investigated in Section~4 for one internal space and in
the Section~5 for two internal spaces. Stable configurations are found for
both of these cases. Some arguments in favour of our Casimir-like ansatz are
listed in the Appendix A. In Appendix B, some general features of a
generalization of the Abel-Plana summation formula to higher dimensional
complex spaces are described. A simple example for a potential with 
nondegenerate minimum at a point with coinciding scale factors is given in
Appendix C.

\section{The model}

\setcounter{equation}{0}
\renewcommand{\theequation}{\thesection\arabic{equation}}

Let us consider a cosmological model with the metric 
\begin{equation}
\label{2.1}ds^2=-e^{2\gamma }d\tau ^2+\sum_{i=0}^ne^{2\beta ^i(\tau
)}g^{(i)}\ , 
\end{equation}
which is defined on the manifold (\ref{1.1}), where the manifolds $M_i$ with
the metrics $g^{(i)}$ are Einstein spaces of dimension $d_i$, i.e. 
\begin{equation}
\label{2.2}R_{m_in_i}\left[ g^{(i)}\right] =\lambda ^ig_{m_in_i}^{(i)}\
,\quad i=0,\ldots ,n 
\end{equation}
and 
\begin{equation}
\label{2.3}R\left[ g^{(i)}\right] =\lambda ^id_i\equiv R_i\ . 
\end{equation}
In the case of constant curvature spaces $\lambda ^i\!=\!k_i\!\left(
d_i-1\right)\!,$\  $k_i=\pm 1,0$. The non-zero components of the Ricci tensor
for the metric (\ref{2.1}) read 
\begin{eqnarray}\label{2.4}
R_{00} & = & -\sum_{i=0}^{n}d_i\left[\ddot \beta ^i - \dot \gamma \dot
\beta + {\left(\dot \beta ^i\right)}^2\right], \\
\nonumber R_{m_in_i} & = & 
g^{(i)}_{m_in_i}\left[\lambda ^i + \exp {\left(2\beta ^i
- 2\gamma\right)}\right.\times \\ 
 &  & \left.\times\left(\ddot \beta ^i + \dot \beta
^i\left(\sum_{i=0}^{n}d_i\dot \beta ^i - \gamma\right)\right)\right] ,
\end{eqnarray}so that the corresponding scalar curvature is given as 
\begin{eqnarray}
\label{2.6}
R & = & \sum_{i=0}^nR_i\exp {\left( -2\beta ^i\right) }+\exp {(-2\gamma
)} \sum_{i=0}^nd_i\times \nonumber \\
 & \times & \left[ 2\ddot \beta ^i-2\dot \gamma \dot \beta ^i+{\left(
\dot \beta ^i\right) }^2+\dot \beta ^i\sum_{j=0}^nd_j\dot \beta ^j\right] . 
\end{eqnarray}
The overdot in formulae (\ref{2.4}) - (\ref{2.6}) denotes differentiation
with respect to time $\tau $.

The action of the model we choose in the following form 
\begin{equation}
\label{2.7}S=\frac 1{2\kappa ^2}\int\limits_Md^Dx\sqrt{|g|}\left\{
R[g]-2\Lambda \right\} +S_c+S_{YGH} , 
\end{equation}
where $D=1+\sum_{i=0}^nd_i$ is the total dimension of the space-time, $%
\Lambda $ is a $D$-dimensional cosmological constant, $\kappa ^2$ --- $D$%
-dimensional gravitational constant and $S_{YGH}$ is the standard
York-Gibbons-Hawking boundary term. The Casimir effect is taken into account
via the additional term $S_c$ of the form \cite{Maeda} 
\begin{equation}
\label{2.8}S_c=-\int\limits_Md^Dx\sqrt{|g|}\rho (x) , 
\end{equation}
where $\rho (x)$ is the Casimir energy density (\ref{A.10}) . For the model
under consideration the Casimir energy is a function of the scale factors
(see (\ref{A.27}) - (\ref{A.29})). After dimensional reduction the action (%
\ref{2.7}) reads 
\begin{equation}
\label{2.9}S=\frac \mu {\kappa ^2}\int d\tau L 
\end{equation}
with Lagrangian $L$ given as 
\begin{eqnarray}\label{2.10}
L & = & \frac{1}{2}e^{-\gamma + \gamma_0}\sum_{i,j=0}^{n}G_{ij}\dot
\beta ^i\dot \beta ^j + \frac{1}{2}e^{\gamma +
\gamma_0}\sum_{i=0}^{n}R_ie^{-2\beta ^i} \nonumber\\
& - & e^{\gamma + \gamma_0}\Lambda -
\frac{\kappa ^2}{\mu}e^{\gamma}F_c .
\end{eqnarray}Here $\gamma _0=\sum_{i=0}^nd_i\beta ^i$ and $\mu
=\prod_{i=0}^n\mu _i$ where $\mu _i$ is defined by the equation (\ref{A.11}%
). The components of the minisuperspace metric read \cite{IMZ} 
\begin{equation}
\label{2.11}G_{ij}=d_i\delta _{ij}-d_id_j\ ,\quad i,j=0,\ldots ,n . 
\end{equation}

It is easy to show that the Euler-Lagrange equations for the Lagrangian (\ref
{2.10}) are equivalent to the Einstein equations for the metric (\ref{2.1})
and the Casimir energy-momentum tensor (\ref{A.13}). This follows from the
fact that for the non-trivial components of the tensor $E_{MN}$ we get 
\begin{eqnarray}\label{2.12}
E_{00} & = & \frac{\partial L}{\partial \gamma}\exp{\left(\gamma -
\gamma _0\right)} , \\
E_{m_in_i} & = &
\frac{1}{d_i}g^{(i)}_{m_in_i}\left(\frac{d}{d\tau}\frac{\partial
L}{\partial \dot \beta ^i} - \frac{\partial L}{\partial \beta
^i}\right)\times \nonumber \\ 
 & & \times\exp {\left(2\beta ^i - \gamma - \gamma _0\right)} ,\\
\nonumber  & & i=0,\ldots ,n\ ,
\end{eqnarray}where the tensor $E_{MN}$ is defined as 
\begin{equation}
\label{2.14}E_{MN}=R_{MN}-\frac 12g_{MN}R-\kappa ^2T_{MN}+\Lambda g_{MN} . 
\end{equation}
For the proof one makes explicit use of equations (\ref{A.10}) and (\ref
{A.12}). %
%%%%%%%%%%%%%%%%%%%%%%%%%%%%%%%%%%%%%%%%%%%%%%%%%%%%%%%%%
%

\section{The effective potential}

\setcounter{equation}{0}
\renewcommand{\theequation}{\thesection\arabic{equation}}

Let us slightly generalize this model to the inhomogeneous case supposing
that the scale factors $\beta ^i=\beta ^i(x)\ (i=0,\ldots ,n)$ are functions
of the coordinates $x$, where $x$ are defined on the $D_0=(1+d_0)$ -
dimensional manifold $\bar M_0=\RR \times M_0$ with the metric 
\begin{equation}
\label{3.1}\bar g^{(0)}=\bar g_{\mu \nu }^{(0)}dx^\mu \otimes dx^\nu
=-e^{2\gamma }d\tau ^2+e^{2\beta ^0(x)}g^{(0)} . 
\end{equation}
$\bar M_0$ denotes the external space-time. The dimensional reduction of the
action (\ref{2.7}) yields 
\begin{eqnarray}\label{3.2}
S & = & \frac{1}{2\kappa ^2_0}\int \limits_{\bar M_0}d^{D_0}x\sqrt {|\bar
g^{(0)}|}\prod_{i=1}^{n}e^{d_i\beta ^i}\left\{R\left[
\bar g^{(0)}\right]\right.- \nonumber \\
 & - & 
G_{ij}\bar g^{(0)\mu \nu}\partial _{\mu}\beta ^i\,\partial _{\nu}\beta
^j +  \sum_{i=1}^{n}R_ie^{-2\beta ^i}- \nonumber \\
 & - & \left.2\Lambda - 2\kappa
^2\rho \right\} ,
\end{eqnarray}where $\kappa _0^2=\kappa ^2/\mu $ is the $D_0$-dimensional
gravitational constant, $\mu =\prod_{i=1}^n\mu _i$ and $G_{ij}\
(i,j=1,\ldots ,n)$ is the midisuperspace metric with the components (\ref
{2.11}). Here the internal scale factors play the role of scalar fields
(dilatons). The action (\ref{3.2}) is written in the Brans-Dicke frame.
Conformal transformation to the Einstein frame 
\begin{equation}
\label{3.3}\hat g_{\mu \nu }^{(0)}={\left( \prod_{i=1}^ne^{d_i\beta
^i}\right) }^{\frac 2{D_0-2}}\bar g_{\mu \nu }^{(0)} 
\end{equation}
yields 
\begin{eqnarray}
\label{3.4}
S & = & \frac 1{2\kappa _0^2}\int\limits_{\bar M_0}d^{D_0}x\sqrt{|\hat
g^{(0)}|}\left\{ \hat R\left[ \hat g^{(0)}\right]\right. \nonumber \\ 
 & - & \left.\bar G_{ij}\hat g^{(0)\mu
\nu }\partial _\mu \beta ^i\,\partial _\nu \beta ^j-2U_{eff}\right\} , 
\end{eqnarray}
where the tensor components of the midisuperspace metric $\bar G_{ij}$ are 
\begin{equation}
\label{3.5}\bar G_{ij}=d_i\delta _{ij}+\frac 1{D_0-2}d_id_j\ ,\quad
i,j=1,\ldots ,n 
\end{equation}
and the effective potential $U_{eff}$ reads 
\begin{eqnarray}
\label{3.6}
U_{eff} & = & {\left( \prod_{i=1}^ne^{d_i\beta ^i}\right) }^{-\frac
2{D_0-2}}\left[ -\frac 12\sum_{i=1}^nR_ie^{-2\beta ^i}+\right.\nonumber\\
 &  & +\left.\Lambda +\kappa
^2\rho \right]  . 
\end{eqnarray}
In what follows we consider the case where the Casimir energy density
depends on the internal scale factors: $\rho =\rho \left( \beta ^1,\ldots
,\beta ^n\right) $. This means that either the external space $M_0$ is
non-compact or the external scale factor is much greater than the internal
ones: $a_0\gg a_1,\ldots ,a_n$.

In the case of one internal space $n=1$ the action and the effective
potential are respectively 
\begin{eqnarray}
\label{3.7}
S & = & \frac 1{2\kappa _0^2}\int\limits_{\bar M_0}d^{D_0}x\sqrt{|\hat
g^{(0)}|}\left\{ \hat R\left[ \hat g^{(0)}\right]- \right.\nonumber\\
 &  & -\left.\hat g^{(0)\mu \nu
}\partial _\mu \varphi \,\partial _\nu \varphi -2U_{eff}\right\} 
\end{eqnarray}
and 
\mathindent=0.5em  
\begin{eqnarray}
\label{3.8}
U_{eff} \!\!\!\!& = & \!\!\!\!
e^{2\varphi {\left[ \frac{d_1}{(D-2)(D_0-2)}\right] }%
^{1/2}}\left[ -\frac 12R_1e^{2\varphi {\left[ \frac{D_0-2}{d_1(D-2)}\right] }%
^{1/2}}\right.+
\!\!\!\!\!\!\!\!\!\!\!\!\!\!\!\!\!\!\!\!\nonumber\\
\!\!\!\! & + & \!\!\!\!\left.\Lambda \kappa ^2\rho (\varphi )\right] , 
\end{eqnarray}
\mathindent=1em
where we redefined the dilaton field as 
\begin{equation}
\label{3.9}\varphi \equiv -\sqrt{\frac{d_1(D-2)}{D_0-2}}\beta ^1\ . 
\end{equation}
The minima of the effective potentials (\ref{3.6}) and (\ref{3.8}) define
stable compactification positions $a_{(c)i}=\exp {\beta _c^i} ,\\
c=1,\ldots ,m$ and small excitations $\psi ^i$ of the internal scale factors
near these positions have a form of minimally coupled massive scalar fields
in the external space-time \cite{GZ}: 
\begin{eqnarray}\label{3.10}
S & = & \frac{1}{2\kappa ^2_0}\int \limits_{\bar M_0}d^{D_0}x\sqrt {|\hat
g^{(0)}|}\left\{\hat R\left[\hat g^{(0)}\right]
- 2\Lambda _{(c)eff}\right\} + \nonumber\\
& + & \sum_{i=1}^{n}\frac{1}{2}\int \limits_{\bar M_0}d^{D_0}x\sqrt {|\hat
g^{(0)}|}\left\{-\hat g^{(0)\mu \nu}\psi ^i_{, \mu}\psi ^i_{, \nu} 
-\right.\nonumber\\
 & -  & \left.m_{(c)i}^2\psi ^i\psi ^i\right\} ,
\end{eqnarray}
where $\Lambda _{(c)eff}\equiv U_{eff}(\vec \beta _c)$ is an
effective cosmological constant and $m_{(c)i}$ are masses of gravitational
excitons corresponding to the $\mbox{\rm c-th}$ minimum. From a physical
point of view it is clear that the effective potential should satisfy
following conditions: 
\begin{eqnarray}\label{}
(i)\; \qquad a_{(c)i} & \mbox{\small $^{>}_{\sim}$}& L_{Pl}\ ,
\nonumber\\
(ii)\ \ \, \quad m_{(c)i} & \leq & M_{Pl}\ ,\nonumber\\
(iii)\quad \Lambda _{(c)eff} & \rightarrow & 0\ .\nonumber
\end{eqnarray}The first condition expresses the fact that the internal
spaces should be unobservable at the present time and stable against quantum
gravitational fluctuations. This condition ensures the applicability of the
classical gravitational equations near positions of minima of the effective
potential. The second condition means that the curvature of the effective
potential should be less than Planckian one. Of course, gravitational
excitons can be excited at the present time if $m_i\ll M_{Pl}$. The third
condition reflects the fact that the cosmological constant at the present
time is very small: $\Lambda \leq 10^{-56}\mbox{cm}^{-2}\approx
10^{-121}\Lambda _{Pl},$ where $\Lambda _{Pl}=L_{Pl}^{-2}$. Thus, for
simplicity, we can demand $\Lambda _{eff}=U_{eff}(\vec \beta _c)=0$. (We
used the abbreviation $\Lambda _{eff}\equiv \Lambda _{(c)eff}$ .) Strictly
speaking, in the multi-minimum case $(c>1)$ we can demand $a_{(c)i}\sim
L_{Pl}$ and $\Lambda _{(c)eff}=0$ only for one of the minima to which
corresponds the present universe state. For all other minima it may be $%
a_{(c)i}\gg L_{Pl}$ and $|\Lambda _{(c)eff}|\gg 0$. %
%%%%%%%%%%%%%%%%%%%%%%%%%%%%%%%%%%%%%%%%%%%%%%%%%%%%%%%%%
%

\section{One internal space}

\setcounter{equation}{0}
\renewcommand{\theequation}{\thesection\arabic{equation}}

Here we consider the case of one internal space $n=1$ or, stricly speaking,
the case where all internal spaces have one common scale factor: $a_1\equiv
a_2\equiv \ldots \equiv a_n\equiv a=\exp \beta $. The latter case
corresponds to the splitting of the internal space into a product of
Einstein spaces \cite{Zh}: $M_1\rightarrow \prod_{i=1}^nM_i$ which leads in
the equations (\ref{3.7}) - (\ref{3.9}) to the substitutions: $%
d_1\rightarrow d=\sum_{i=1}^nd_i\ ;\quad R_1\rightarrow R=\sum_{i=1}^nR_i$.
For effective one-scale-factor models the Casimir energy density reads
according to (\ref{A.25}): $\rho =C\exp {(-D\beta )}$, where $C$ is a
constant that strongly depends on the topology of the model. For example,
for fluctuations of massless scalar fields the constant $C$ was calculated
to take the values: $C=-8.047\cdot 10^{-6}$ if $\bar M_0=\RR \times
S^3\ ,\ M_1=S^1$ (with $e^{\beta ^0}$ as scale factor of $S^3$ and $e^{\beta
^0}\gg e^{\beta ^1}$) \cite{Koik}; $C=-1.097$ if $\bar M_0=\RR \times 
\RR ^2\ ,\ M_1=S^1$ \cite{DeW} and $C=3.834\cdot 10^{-6}$ if $\bar M_0=%
\RR \times S^3\ ,\ M_1=S^3$ (with $e^{\beta ^0}\gg e^{\beta ^1}$) 
\cite{Koik}.

For an effective potential with $\rho =C\exp {(-D\beta )}$ the
zero-extremum-conditions $\left. \frac{\partial U_{eff}}{\partial \beta }%
\right| _{min}=0$ and $\Lambda _{eff}=0$ lead to a fine tuning of the
parameters of the model 
\begin{equation}
\label{5.2}R_1e^{-2\beta _c}=\frac{2D}{D-2}\Lambda ,\quad R_1e^{(D-2)\beta
_c}=\kappa ^2CD,\ 
\end{equation}
which implies $\mbox{\rm sign}\,R_1=\mbox{\rm sign}\,\Lambda =\mbox{\rm sign}%
\,C$. We note that a similar fine tuning was obtained by different methods
in papers \cite{Maeda} (for one internal space) and \cite{BZ1} (for $n$
identical internal spaces). To get a minimum, the second derivative of the
effective potential at the extremum should be positive: 
\begin{equation}
\label{4.3}{\left. \frac{\partial ^2U_{eff}}{\partial \beta ^2}\right| }%
_{\beta _c}=(D-2)R{_1\left( e^{-2\beta _c}\right) }^{\frac{D-2}{D_0-2}}>0. 
\end{equation}
Thus, a stable compactification takes place if the internal space has a
positive scalar curvature $R_1>0$ (or for a split space $M_1$ the sum of the
curvatures of the constituent spaces $M_1^k$ should be positive). The mass
squared of the gravitational excitons is given by the expression 
\begin{equation}
\label{4.4}m^2\equiv {\left. \frac{\partial ^2U_{eff}}{\partial \varphi ^2}%
\right| }_{\varphi _c}=\frac{D_0-2}{d_1}R{_1\left( e^{-2\beta _c}\right) }^{
\frac{D-2}{D_0-2}}. 
\end{equation}

As example, let us consider a manifold $M$ with topology $M=\RR \times
S^3\times S^3$ where $a_0\gg a_1$. Then according to \cite{Koik} the
constant $C$ is given as $C=3.834\cdot 10^{-6}>0,$ so that with $%
C,R_1=d_1(d_1-1)=6>0$ the effective potential has a minimum provided $%
\Lambda >0$. Normalizing $\kappa _0^2$ to unity, we get $\kappa ^2=\mu ,$
where $\mu =\left. 2\pi ^{(d+1)/2}\right/ \Gamma \left( \frac 12(d+1)\right) 
$ is the volume of the $d$-dimensional sphere. For the model with the
topology $M=\RR \times S^3\times S^3$ we obtain $a_c\approx 1.5\cdot
10^{-1}L_{Pl}$ and $m\approx 2.12\cdot 10^2M_{Pl}$. Hence, for this
particular example the conditions (i) and (ii) are not satisfied . It would
be interesting to investigate models with more complex topology. Of course,
the calculation of the Casimir effect in this case becomes a very
complicated problem. Therefore in the Appendix A we propose ad hoc the
Casimir-like expressions (\ref{A.27}), (\ref{A.28}) for the energy density
in the case of more than one internal spaces with non-identical scale
factors. In the next section we investigate the stable compactification
problem for this potential in the case of two internal spaces. For this case
the Casimir-like potential considerably simplifies, but even in this case
the stability analysis is still complicated. To our knowledge the
investigation of the stable compactification problem for models with more
than one non-identical (with respect to the scale factors) spaces was not
performed up to now. %
%%%%%%%%%%%%%%%%%%%%%%%%%%%%%%%%%%%%%%%%%%%%%%%%%%%%%%%%%
%

\section{Two internal spaces}

\setcounter{equation}{0}
\renewcommand{\theequation}{\thesection\arabic{equation}}

After these brief considerations on Casimir potentials for one-scale-factor
models we turn now to some methods applicable for an analysis of
two-scale-factor models with Casimir-like potentials. Some arguments in
favour of such potentials are given in Appendix A. We propose to use these
potentials of the general form 
\begin{eqnarray}
\label{5.4}
\rho & = & e^{-\sum_{i=0}^nd_i\beta ^i}\sum_{k_0,\ldots
,k_n=0}^n\epsilon _{k_0|k_1\ldots k_n|}\sum_{\xi _0=0}^{d_{k_1}}\sum_{\xi
_1=0}^{d_{k_2}}\ldots 
\!\!\!\!\!\!\!\!\!\!\!\!\!\!\!\!\!\!\!\!\!\nonumber\\
 & \ldots & \sum_{\xi _{n-1}=0}^{d_{k_n}}A_{\xi _0\ldots \xi
_{n-1}}^{(k_0)}\frac{{\left( e^{\beta ^{k_1}}\right) }^{\xi _0}\ldots {%
\left( e^{\beta ^{k_n}}\right) }^{\xi _{n-1}}}{{\left( e^{\beta
^{k_0}}\right) }^{\xi _0+\xi _1+\cdots +\xi _{n-1}+1}}\ , 
\!\!\!\!\!\!\!\!\!\!\!\!\!\!\!\!\!\!\!\!\!\nonumber\\
 & & 
\end{eqnarray}
in order to achieve a first crude insight into a possible stabilization
mechanism of internal space configurations due to exact Casimir potentials
depending on $n$ scale factors. In (\ref{5.4}) 
$A_{\xi _0\ldots \xi _{n-1}}^{(k_0)}$ are dimensionless constants
which depend on the topology of the model and 
$\epsilon _{ik\ldots m}={(-1)}%
^P\varepsilon _{ik\ldots m}$. Here 
$\varepsilon _{ik\ldots m}$ 
is the totally antisymmetric symbol
($\varepsilon _{01\ldots m}=+1$), $P$ is the number of the 
permutations of the $%
01\ldots n$ resulting in $ik\ldots m$. $|k_1k_2\ldots k_n|$ means summation
is taken over $k_1<k_2<\cdots <k_n$.

{From} investigations performed in the last decades (see e.g. \cite{MT2,Eliz},
Refs. therein and Appendix A below) we know that exact Casimir potentials
can be expressed in terms of Epstein zeta function series with scale factors
as parameters. Unfortunately, the existing representations of these zeta
function series are not well suited for a stability analysis of the
effective potential $U_{eff}$ as function over the total target space $\vec
\beta \in $ $\RR
_T^n$. The problems can be circumvented partially by the use of asymptotic
expansions of the zeta function series in terms of elementary functions for
special subdomains $\Omega _a$ of the target space $\Omega _a$ $\subset \RR
_T^n$. According to \cite{MT2,Eliz} potential (\ref{5.4}) gives a crude
approximation of exact Casimir potentials in subdomains $\Omega _a$. In
contrast with other approximative potentials proposed in literature \cite
{AGHK,HKVW} potential (\ref{5.4}) shows a physically correct behavior under
decompactification of factor-space components. The question, in as far (\ref
{5.4}) can be used in regions $\RR
_T^n\setminus \Omega _a$, needs an additional investigation. The philosophy
of the proposed method consists in a consideration of potentials (\ref{5.4})
on the whole target space $\RR
_T^n$, and testing of scale factors and parameters of possible minima of the
corresponding effective potential on their compatibility with asymptotic
approximations of exact Casimir potentials in $\Omega _a$. As a beginning,
we describe in the following only some techniques, without explicit
calculation and estimation of exciton masses.

Before we start our analysis of two-scale-factor models with Casimir-like
potentials 
\begin{eqnarray}
\label{5.5}
\rho & = & e^{-\sum_{i=1}^2d_i\beta ^i}\left[
\sum_{i=0}^{d_2}A_i^{(1)} \frac{{e^{i\beta ^2}}}{{e^{(D_0+i)\beta ^1}}}%
+\right.\nonumber\\
 &  & +\left.\sum_{j=0}^{d_1}A_j^{(2)}%
 \frac{{e^{j\beta ^1}}}{{e^{{(D_0+j)}\beta ^2}}}%
\right] 
\end{eqnarray}
let us introduce the following convenient (temporary) notations: 
\mbox{$x:=a_1\equiv \ {e^{\beta ^1}}$,}\ $y:=a_2\equiv {e^{\beta ^2}},\
P_\xi :=\kappa ^2A_\xi ^{(1)},\ S_\xi :=\kappa ^2A_\xi ^{(2)}.\ $In terms of
these notations the effective potential (\ref{3.6}) reads 
\begin{eqnarray} 
\label{5.6}  
U_{eff} & = & (x^{d_1}y^{d_2})^{-\frac 2{D_0-2}}\left[ - 
\frac{R_1}2x^{-2}-\frac{R_2}2y^{-2}+\Lambda +\right. \nonumber\\  
 & & +x^{-d_1}y^{-d_2}\left(
\sum_{i=0}^{d_2}P_iy^ix^{-(D_0+i)}\right. + \nonumber\\
 &  & \left.\left. + \sum_{j=0}^{d_1}S_jx^jy^{-(D_0+j)}\right)
\right] . 
\end{eqnarray}
For physically relevant configurations with scale factors near Planck length 
\begin{equation}
\label{5.7}0<x,y<\infty 
\end{equation}
we transform extremum conditions $\partial _{\beta
^{1,2}}U_{eff}=0\Leftrightarrow \partial _{x,y}U_{eff}=0$ by factoring out
of $(xy)^{-D}-$terms and taking combinations 
\mbox{$\partial _xU_{eff}\pm \partial _yU_{eff}=0$} to an equivalent system
of two algebraic equations in $x$ and $y$ : 
\vspace{-1ex}
\mathindent=-1em  
\begin{eqnarray}
\label{5.8} 
 & & I_{1+}  =  (xy)^{D-2}\left[ 
\frac{D-2}{D_0-2}\left( R_1y^2+R_2x^2\right) -\right.
\nonumber\!\!\!\!\!\!\!\!\!\\
 & & -  \left.\frac{2\Lambda }{D_0-2}%
D^{^{\prime }}x^2y^2\right] - \left( 
\frac{2D^{^{\prime }}}{D_0-2}+D\right)\times 
\nonumber\!\!\!\!\!\!\!\!\!\\
 & & \times  \left[
\sum_{i=0}^{d_2}P_iy^{D_0+d_1+i}x^{d_2-i}+
 \sum_{j=0}^{d_1}S_jy^{d_1-j}x^{D_0+d_2+j}\right] 
\nonumber\!\!\!\!\!\!\!\!\!\\  
 & & =  0 \nonumber\!\!\!\!\!\!\!\!\!\\  
&  &  \nonumber\!\!\!\!\!\!\!\!\!\\ 
 & & I_{1-}  =  (xy)^{D-2}\left[ 
\frac{d_1-d_2}{D_0-2}\left( R_1y^2+R_2x^2\right) +\right. 
\nonumber\!\!\!\!\!\!\!\!\!\\
 & & +  \left.\left(R_1y^2-R_2x^2\right) -   
\frac{2\Lambda }{D_0-2}\left( d_1-d_2\right) x^2y^2\right] - 
\nonumber\!\!\!\!\!\!\!\!\!\\  
 & & -  \sum_{i=0}^{d_2}P_i\left[ D_0\left( 
\frac{d_1-d_2}{D_0-2}+1\right) +2i\right] y^{D_0+d_1+i}x^{d_2-i}- 
\nonumber\!\!\!\!\!\!\!\!\!\\  
 & & -  \sum_{j=0}^{d_1}S_j\left[ D_0\left( 
\frac{d_1-d_2}{D_0-2}-1\right) -2j\right] y^{d_1-j}x^{D_0+d_2+j} 
\nonumber\!\!\!\!\!\!\!\!\!\\  
 & & =  0.\!\!\!\!\!\!\!\!\! 
\end{eqnarray}
(Here we used the notation $D^{^{\prime }}:=d_1+d_2.$) Thus, scale factors $%
a_1$ and $a_2$ satisfying the extremum conditions are defined as common
roots of polynomials (\ref{5.8}). %, (\ref{5.8a}).
In the general case of arbitrary dimensions $(D_0,\ d_1,\ d_2)$ and
arbitrary parameters $\left\{ R_1,R_2,P_i,S_i\right\} $ these roots are
complex, so that only a restricted subclass of them are real and fulfill
condition (\ref{5.7}). In the following we derive necessary conditions on
the parameter set guaranteeing the existence of real roots satisfying (\ref
{5.7}). The analysis could be carried out using resultant techniques \cite
{34} on variables $x,\ y$ directly. The structure of $I_{1\pm }$ suggests
another, more convenient method \cite{35}. Introducing the projective
coordinate $\lambda =y/x$ we rewrite (\ref{5.8}) %, (\ref{5.8a})
as $I_{1\pm }=x^DI_{2\pm }(y,\lambda )$ with 
\mathindent=-0.5em
\begin{equation}
\label{5.9} 
\begin{array}{lr}
I_{2+}  =  -a_0(\lambda )+a_{D-2}(\lambda )y^{D-2}-y^D\Delta _{+}  =  0
 & \!\!\!(a) \!\!\!\!\!\!\!\!\\  
  &  \!\!\!\!\!\!\!\!\\ 
I_{2-}  =  -b_0(\lambda )+b_{D-2}(\lambda )y^{D-2}-y^D\Delta _{-}  =  0
 & \!\!\!(b) \!\!\!\!\!\!\!\!
\end{array}
\end{equation}
\mathindent=-2em 
and coefficient-functions 
\begin{eqnarray}
\label{5.10} 
 & & a_0(\lambda )  =  [ 
\frac{2D^{^{\prime }}}{D_0-2}+D]\left[ \sum_{i=0}^{d_2}P_i\lambda
^{D_0+d_1+i}+\sum_{j=0}^{d_1}S_j\lambda ^{d_1-j}\right] \nonumber\!\!\!\!\!\\  
& & 
a_{D-2}(\lambda )  =  \frac{D-2}{D_0-2}\left( R_1\lambda ^2+R_2\right)   
\nonumber\!\!\!\!\!\\ 
 & & b_0(\lambda )  =  \sum_{i=0}^{d_2}P_i\left[ D_0\left( 
\frac{d_1-d_2}{D_0-2}+1\right) +2i\right] \lambda ^{D_0+d_1+i}+ 
\nonumber\!\!\!\!\!\\  
 & & \quad \quad \quad +\sum_{j=0}^{d_1}S_j\left[ D_0\left( 
\frac{d_1-d_2}{D_0-2}-1\right) -2j\right] \lambda ^{d_1-j} 
\nonumber\!\!\!\!\!\\  
 & & b_{D-2}(\lambda )=  \frac{d_1-d_2}{D_0-2}\left( R_1\lambda ^2+R_2\right)
+\left( R_1\lambda ^2-R_2\right) \nonumber\!\!\!\!\!\\  
 & & \Delta _{\pm } =  \frac{2\Lambda }{D_0-2}(d_1\pm d_2). \!\!\!\!\!
\end{eqnarray}
\mathindent=1em 
Equations (\ref{5.9}) have common roots if the coefficient functions $%
\{a_i(\lambda ),b_i(\lambda )\}$ are connected by a constraint. This
constraint is given by the vanishing resultant 
\begin{equation}
\label{5.11}R_y[I_{2+},I_{2-}]=w(\lambda )=0. 
\end{equation}
Now, the roots can be obtained in two steps. First, one finds the set of
roots $\{\lambda _i\}$ of the polynomial $w(\lambda )$. Physical condition (%
\ref{5.7}) on the affine coordinates $(x,y)$ implies here a corresponding
condition on the projective coordinate $\lambda =y/x$ 
\begin{equation}
\label{5.12}Im(\lambda )=0,\qquad 0<\lambda <\infty . 
\end{equation}
Second, one searches for each $\lambda _i$ solutions $\{y_{ij}\}$ of (\ref
{5.9}). The complete set of physically relevant solutions of system (\ref
{5.8}) %(\ref{5.8a})
is then given in terms of pairs $\{x_{ij}=y_{ij}/\lambda _i,\ y_{ij}\}$.

Because of the simple $y-$structure of equations (\ref{5.9}) the polynomial $%
w(\lambda )$ can be derived from (\ref{5.9}) directly, without explicit
calculation of the resultant. Taking $b_0(\lambda )I_{2+}-a_0(\lambda
)I_{2-}=0$, \mbox{$\Delta _{-}I_{2+}-\Delta _{+}I_{2-}=0$} and assuming $y>0$
we get 
\begin{equation}
\label{5.13}y^2=\frac{L_3}{L_1},\qquad y^{D-2}=\frac{L_1}{L_2}, 
\end{equation}
where 
\begin{eqnarray}
\label{5.14} 
L_1(\lambda ) & := & \Delta _{-}a_0(\lambda )-\Delta _{+}b_0(\lambda ) 
\nonumber\\  
L_2(\lambda ) & := & \Delta _{-}a_{D-2}(\lambda )-\Delta _{+}b_{D-2}(\lambda
) \nonumber\\  
L_3(\lambda ) & := & a_0(\lambda )b_{D-2}(\lambda )-b_0(\lambda
)a_{D-2}(\lambda )  
\end{eqnarray}
depend only on $\lambda $. Excluding $y$ from (\ref{5.13}) yields the
necessary constraint for the coefficient functions of equation system (\ref
{5.9}) 
\begin{equation}
\label{5.15}w(\lambda )=L_2^2(\lambda )L_3^{D-2}(\lambda )-L_1^D(\lambda
)=0. 
\end{equation}
Together with condition (\ref{5.12}), this polynomial of degree 
\begin{equation}
\deg{}_\lambda [w(\lambda )]=D^2 
\end{equation}
can be used for a first test of internal space configurations on stability
of their compactification. If the corresponding parameters $\left\{
R_1,R_2,P_i,S_i\right\} $ allow the existence of positive real roots $%
\lambda _i$, the space configuration is a possible candidate for a stable
compactified configuration and can be further tested on the existence of
minima of the effective potential $U_{eff}.$ Otherwise it belongs to the
class of unstable internal space configurations.

Before we turn to the consideration of two-scale-factor models with factor-%
spaces of the same topological type $(M_1=M_2)$ we note that for the
coefficient functions (\ref{5.14}), because of (\ref{5.7}) and (\ref{5.13}),
there must hold 
\begin{equation}
\label{5.16}\mbox{\rm sign}(L_1)\big|
_{\lambda _i}=\mbox{\rm sign}(L_2)\big|
_{\lambda _i}=\mbox{\rm sign}(L_3)\big|
_{\lambda _i}. 
\end{equation}
Furthermore we see from (\ref{5.15}) that for even dimensions  $D=\dim
(M_1)+\dim (M_2)+\dim (M_0)+1$ of the product-manifold the polynomial $%
w(\lambda )$ factors into two subpolynomials of degree $D^2/2$%
\begin{eqnarray}
\label{5.17}
w(\lambda ) & = & \left[ L_2(\lambda )L_3^{\frac{D-2}2}(\lambda
)+L_1^{\frac D2}(\lambda )\right]\times\nonumber\\ 
 &  & \times\left[ L_2(\lambda )L_3^{\frac{D-2}%
2}(\lambda )-L_1^{\frac D2}(\lambda )\right] \nonumber\\
 & = &0. 
\end{eqnarray}

\subsection{ Two identical internal factor-spaces\label{mark1}}

In the case of identical internal factor-spaces $M_1$ and $M_2$ we have $\ d{%
_1=d_2,P_i=S_i,\ R_1=R_2}$. If we assume additionally an external space-time 
$\bar M_0$ with $\dim \bar M_0=4$ and, hence, ${\ D=2(d_1+2),\ }$ then
equations (\ref{5.9}) and polynomial (\ref{5.17}) can be rewritten as 
\mathindent=0em 
\begin{equation}
\label{5.18} 
\begin{array}{lll}
I_{2+}= & -4(d_1+1)\bar a_0(\lambda )+(d_1+1)R_1(\lambda ^2+1)y^{D-2}- 
\!\!\!\!\!\!\!\!\!\!\!\!\!\!\!\!\!\!\!\! 
& \!\!\!\!\!\!\!\!\!\!\!\!\!\!\!\!\!\!\!\!  %\nonumber
\\  
 & 
\!\!\!\!\!\!\!\!\!\!\!\!\!\!\!\!\!\!\!\! 
& \!\!\!\!\!\!\!\!\!\!\!\!\!\!\!\!\!\!\!\! \\  
 & -2d_1\Lambda y^D=0 
\!\!\!\!\!\!\!\!\!\!\!\!\!\!\!\!\!\!\!\! 
 &  \!\!\!\!\!\!\!\! (a) \!\!\!\!\!\!\!\!\!\!\!\!%\nonumber
 \\  
 & 
\!\!\!\!\!\!\!\!\!\!\!\!\!\!\!\!\!\!\!\! 
& \!\!\!\!\!\!\!\!\!\!\!\!\!\!\!\!\!\!\!\! \\ 
I_{2-}= & (\lambda ^2-1)\stackunder{\bar I_{2-}}{\underbrace{\left[ -2\bar
b_0(\lambda )+R_1y^{D-2}\right] }}=0 
\!\!\!\!\!\!\!\!\!\!\!\!\!\!\!\!\!\!\!\! 
& \!\!\!\!\!\!\!\! (b) \!\!\!\!\!\!\!\!\!\!\!\!
\end{array}
\end{equation}
and 
\mathindent=-2em 
\begin{eqnarray}
\label{5.19} 
 & & w(\lambda ) =4\Lambda ^2d_1^2\left[ 2(\lambda ^2-1)\right]
^{2(d_1+2)}\times \nonumber\!\!\!\!\!\!\!\!\!\!\!\!\\  
& & \times 
\stackunder{w_{+}(\lambda )}{\underbrace{\left[ R_1^{d_1+2}\left(
2(d_1+1)\right) ^{d_1+1}\bar L_3^{d_1+1}+\left( -2\Lambda d_1\right)
^{d_1+1}\bar b_0^{d_1+2}\right] }} \nonumber\!\!\!\!\!\!\!\!\!\!\!\!\\  
 & & \times 
\stackunder{w_{-}(\lambda )}{\underbrace{\left[ R_1^{d_1+2}\left(
2(d_1+1)\right) ^{d_1+1}\bar L_3^{d_1+1}-\left( -2\Lambda d_1\right)
^{d_1+1}\bar b_0^{d_1+2}\right] }} \nonumber\!\!\!\!\!\!\!\!\!\!\!\!\\  
 & & =0  \!\!\!\!\!\!\!\!\!\!\!\!
\end{eqnarray}
\mathindent=1em 
with the notations 
\begin{eqnarray}
\label{5.20} 
\bar L_3 & := & 2\bar a_0-(\lambda ^2+1)\bar b_0 \nonumber\\  
\bar a_0 & := & \sum_{i=0}^{d_1}P_i\left[ \lambda ^{4+d_1+i}+\lambda
^{d_1-i}\right] \nonumber\\  
\bar b_0 & := & \sum_{i=0}^{d_1}P_i(2+i)\lambda
^{d_1-i}\sum_{j=0}^{i+1}\lambda ^{2j}. 
\end{eqnarray}
According to the factorizing polynomial (\ref{5.18}(b)) and the constraint (%
\ref{5.19}) the root set splits into two types of subsets defined by the
conditions 
\mathindent=-1em
\begin{eqnarray}
\label{5.20a} 
 & & \mbox{\rm type I:} \nonumber\!\!\!\!\!\!\!\!\!\!\!\! \\
 & & I_{2+}(\lambda ,y)=0,\quad \bar I_{2-}(\lambda ,y)=0,\quad w_{+}(\lambda
)w_{-}(\lambda )=0  \nonumber\!\!\!\!\!\!\!\!\!\!\!\!\\  
&  &  \nonumber\!\!\!\!\!\!\!\!\!\!\!\!\\[-1ex] 
 & & \mbox{\rm type II:}\nonumber\!\!\!\!\!\!\!\!\!\!\!\!\\
 & & I_{2+}(\lambda ,y)=0,\quad \lambda ^2-1=0 . 
 \!\!\!\!\!\!\!\!\!\!\!\!
\end{eqnarray}
\mathindent=1em
For roots of type I we have similar to (\ref{5.13}) 
\begin{equation}
\label{5.21}y^2=-\frac{(d_1+1)R_1\bar L_3}{2\Lambda d_1\bar b_0},\qquad
y^{D-2}=\frac{2\bar b_0}{R_1}, 
\end{equation}
whereas type II roots at $\lambda =1$ should be found from the polynomial $%
I_{2+}$ (\ref{5.18}(a)) directly. For this polynomial we have now simply 
\mathindent=0.5em
\begin{eqnarray}
\label{5.22}
I_{2+}(\lambda =1)\!\!\!\! & := &\!\!\!\! \frac{\Lambda d_1}{d_1+1}%
y^{2(d_1+2)}-R_1y^{2(d_1+1)}+4\sum_{i=0}^{d_1}P_i 
\nonumber\!\!\!\!\!\!\!\!\!\!\!\!\\
 \!\!\!\! & = & \!\!\!\!0. \!\!\!\!\!\!\!\!\!\!\!\!
\end{eqnarray}
Coming back to the general case of identical factor-spaces $M_1,\ M_2$ with
coinciding or noncoinciding scale factors we note that there exists an
interchange symmetry between $M_1$ and $M_2$, which becomes apparent in the
root structure of the polynomial $w(\lambda )$. From 
\begin{eqnarray}
\label{5.23} 
U_{eff}\!\!\!\! & = & \!\!\!\!(xy)^{-d_1}\left[ - 
\frac{R_1}2(x^{-2}-y^{-2})+\Lambda +\right. \nonumber\\  
 \!\!\!\! & + & \!\!\!\!\left. (xy)^{-d_1}\sum_{i=0}^{d_1}P_i\left(
y^ix^{-4-i}+x^iy^{-4-i}\right) \right] 
\end{eqnarray}
\mathindent=1em 
we see that $x$ and $y$ enter (\ref{5.23}) symmetrically. When one extremum
of (\ref{5.23}) is located at $\{x_i=a,\ y_i=b\}$ then because of the
interchange symmetry $x\rightleftharpoons y$ there exists a second extremum
located at $\{x_j=b,\ y_j=a\}$. So we have for the corresponding projective
coordinates : 
\mathindent=-1em
\begin{eqnarray}
\label{5.24}
 & & \lambda _i=y_i/x_i=b/a,\quad \lambda _j=y_j/x_j=a/b\nonumber\!\!\!\!\!\!\!\!\!\\
 & & \Longrightarrow \lambda _i=\lambda _j^{-1}.\!\!\!\!\!\!\!\!\! 
\end{eqnarray}
\mathindent=1em
By regrouping of terms in (\ref{5.19}) it is easy to show that 
\begin{equation}
\label{5.25}w(\lambda ^{-1})=\lambda ^{-D^2}w(\lambda ) 
\end{equation}
and, hence, roots $\{\lambda _i\neq 0\}$ of $w(\lambda )=0$ exist indeed in
pairs $\{\lambda _i,\ \lambda _i^{-1}\}.$ But there is no relation
connecting this root-structure with a symmetry between $w_{+}(\lambda )$ and 
$w_{-}(\lambda )$ in (\ref{5.19}) $w_{+}(\lambda ^{-1})\not \sim
w_{-}(\lambda )$. For completeness, we note that relation (\ref{5.24}) is
formally similar to dualities recently investigated in superstring theory 
\cite{36a}.

Before we turn to an analysis of minimum conditions for effective potentials 
$U_{eff}$ corresponding to special classes of solutions of $w(\lambda )=0$
we rewrite the necessary second derivatives 
\begin{eqnarray}
\label{5.25a} 
   \partial _{xx}^2U_{eff} & = 
  & -\frac{R_1}2\left( \alpha _1x^{-d_1-4}y^{-d_1}+\right. 
\nonumber\!\!\!\!\!\!\!\!\!\\
 & + &  \left.\alpha
_2x^{-d_1-2}y^{-d_1-2}\right) + \nonumber\!\!\!\!\!\!\!\!\!\\  
  & + & \Lambda \alpha _2x^{-d_1-2}y^{-d_1}+ \nonumber\!\!\!\!\!\!\!\!\!\\  
  & + &  \sum_{i=0}^{d_1}P_i\left( \alpha _3y^{i-2d_1}x^{-i-2d_1-6}\right.+
\nonumber\!\!\!\!\!\!\!\!\!\\
 & + & \left.\alpha
_4x^{i-2d_1-2}y^{-i-2d_1-4}\right) \nonumber\!\!\!\!\!\!\!\!\!\\[2ex]
  \partial _{yy}^2U_{eff} & = & \partial _{xx}^2U_{eff} 
\Big|_{x\rightleftharpoons y} \nonumber\!\!\!\!\!\!\!\!\!\\[2ex]  
 \partial _{xy}^2U_{eff} &  = 
 & - 
\frac{R_1}2\alpha _5\left( x^{-d_1-3}y^{-d_1-1}\right.+ 
\nonumber\!\!\!\!\!\!\!\!\!\\
 & + & \left.x^{-d_1-1}y^{-d_1-3}\right)
+ \nonumber\!\!\!\!\!\!\!\!\!\\  
&  + & \Lambda \alpha _6x^{-d_1-1}y^{-d_1-1}+\nonumber\!\!\!\!\!\!\!\!\!\\
 & + & \sum_{i=0}^{d_1}P_i\alpha _7\left(y^{i-2d_1-1}x^{-i-2d_1-5}\right.+
 \nonumber\!\!\!\!\!\!\!\!\!\\
 & + & \left.\alpha _4x^{i-2d_1-1}y^{-i-2d_1-5}\right),
\end{eqnarray}
\mathindent=1em
where 
\begin{eqnarray}
\label{5.25b} 
\alpha _1 & = & (d_1+2)(d_1+3) \nonumber\\ 
\alpha _2 & = & d_1(d_1+1) \nonumber\\  
\alpha _3 & = & (2d_1+i+4)(2d_1+i+5) \nonumber\\
\alpha _4 & = & (2d_1-i)(2d_1-i+1) \nonumber\\
\alpha _5 & = & d_1(d_1+2) \nonumber\\
\alpha _6 & = & d_1^2 \nonumber\\
\alpha _7 & = & (2d_1+i+4)(2d_1-i), 
\end{eqnarray}
in the more appropriate form (notation $\widetilde{\mu }=\lambda
^{d_1}y^{-2D+2}$) 
\begin{eqnarray}
\label{5.25c} 
\partial _{xx}^2U_{eff} & = & \lambda ^2 
\widetilde{\mu }\left[ -\frac{R_1}2\left( \alpha _1\lambda ^2+\alpha
_2\right) y^{D-2}+\Lambda \alpha _2y^D\right. 
\!\!\!\!\!\!\!\!\!\!\!\!\!\!\!\!\!\nonumber\\  
 & + & \left. \sum_{i=0}^{d_1}P_i\left( \alpha _3\lambda ^{4+d_1+i}+\alpha
_4\lambda ^{d_1-i}\right) \right] 
\!\!\!\!\!\!\!\!\!\!\!\!\!\!\!\!\!\nonumber\\[2ex]  
\partial _{yy}^2U_{eff} & = & 
\widetilde{\mu }\left[ -\frac{R_1}2\left( \alpha
_1+\alpha _2\lambda ^2\right) y^{D-2}+\Lambda \alpha _2y^D\right. 
\!\!\!\!\!\!\!\!\!\!\!\!\!\!\!\!\!\nonumber\\  
 & + & \left.\sum_{i=0}^{d_1}P_i\left( \alpha _4\lambda ^{4+d_1+i}+\alpha
_3\lambda ^{d_1-i}\right) \right] 
\!\!\!\!\!\!\!\!\!\!\!\!\!\!\!\!\!\nonumber\\[2ex]  
\partial _{xy}^2U_{eff} & = & \lambda 
\widetilde{\mu }\left[ -\frac{R_1}2\alpha _5\left( \lambda ^2+1\right)
y^{D-2}+\Lambda \alpha _6y^D\right. 
\!\!\!\!\!\!\!\!\!\!\!\!\!\!\!\!\!\nonumber\\  
 & + & \left. \sum_{i=0}^{d_1}P_i\alpha _7\left( \lambda ^{4+d_1+i}+\lambda
^{d_1-i}\right) \right] .  \!\!\!\!\!\!\!\!\!\!\!\!\!\!\!\!\!\nonumber\\
 & & \!\!\!\!\!\!\!\!\!\!\!\!\!\!\!\!\!
\end{eqnarray}
Introducing the notations 
\begin{equation}
\label{5.25c001}\widetilde{A}_c:=\left( 
\begin{array}{cc}
\partial _{xx}^2U_{eff} & \partial _{xy}^2U_{eff} \\ 
\partial _{xy}^2U_{eff} & \partial _{yy}^2U_{eff} 
\end{array}
\right) 
\end{equation}
and 
\mathindent=0.2em
\begin{equation}
\label{5.25c002}w_{(c)1,2}:=\frac 12\left[ Tr(\widetilde{A}_c)\pm \sqrt{%
Tr^2( \widetilde{A}_c)-4\det (\widetilde{A}_c)}\right] 
\end{equation}
\mathindent=1em
the minimum conditions are given as 
\begin{equation}
\label{5.25c003}w_{(c)1}>0,\ w_{(c)2}\geq 0. 
\end{equation}
In the degenerate case of coinciding scale factors $x=y,\ \lambda =1$ there
hold the following relations between the derivatives of effective potentials 
$U_{eff}(x,y)$ and $\widetilde{U}_{eff}(y)=U_{eff}(y,y)$%
\begin{eqnarray}
\label{5.25c01} 
\partial _y 
\widetilde{U}_{eff} & = & 
\partial _xU_{eff}\big| _{x=y}+\partial _yU_{eff}\big| %
_{x=y} \nonumber\\   
\partial _{yy}^2\widetilde{U}_{eff} & = & \partial _{xx}^2U_{eff}\big| %
_{x=y}+\partial _{yy}^2U_{eff}\big| _{x=y} \nonumber\\
 & + & 2\partial _{xy}^2U_{eff}\big| %
_{x=y} 
\end{eqnarray}
and minimum conditions reduce to 
\begin{equation}
\label{5.25c02}\partial _y\widetilde{U}_{eff}=0,\quad \partial _{yy}^2 
\widetilde{U}_{eff}>0 
\end{equation}
with 
\mathindent=1em
\begin{eqnarray}
\label{5.25c03} 
\partial _{yy}^2\widetilde{U}_{eff} & = & 2y^{-2D+2}\left[
-R_1(d_1+1)(2d_1+3)y^{D-2}\right. \!\!\!\!\!\!\!\!\nonumber\\  
 & + & \Lambda d_1(2d_1+1)y^D \!\!\!\!\!\!\!\nonumber\\
 & + & \left.4(d_1+1)(4d_1+5)\sum_{i=0}^{d_1}P_i\right]. \!\!\!\!\!\!\!\!
\end{eqnarray}
\mathindent=1em
For convenience of the additional explicit calculations of constraint $%
U_{eff}\big|
_{\min }=0$ we rewrite also effective potential (\ref{5.23}) in terms of
variables $y,\ \lambda $%
\begin{eqnarray}
\label{5.25c04} 
U_{eff} & = & \lambda ^{d_1}y^{-2D+4}\left[ - 
\frac{R_1}2y^{D-2}(\lambda ^2+1)+\Lambda y^D+\right. 
\!\!\!\!\!\!\!\nonumber\\ 
 & + & \left. \sum_{i=0}^{d_1}P_i\left( \lambda ^{4+d_1+i}+\lambda
^{d_1-i}\right) \right] . \!\!\!\!\!\!\!
\end{eqnarray}
The further analysis consists in a compatibility consideration of minimum
conditions (\ref{5.25c003}) and (\ref{5.25c02}) with properties of the
polynomial $w(\lambda )$, expressions like (\ref{5.21}) defining $y^{D-2}$
and $y^D=y^{D-2}y^2$ as functions of $\lambda $ on the parameter-space 
$\RR _{par}^{d_1+3}=\{(R_1,\Lambda ,P_i)\mid i=0,\ldots ,d_1\}$ and
the constraint $U_{eff}\big|
_{\min }=0$. As result we will get a first crude division of $\RR %
_{par}^{d_1+3}$ in stability-domains allowing the existence of minima of the
effective potential $U_{eff}$ and forbidden regions corresponding to
unstable internal space configurations.

After these general considerations we turn now to a more concrete analysis.

\subsection{Noncoinciding scale factors $(\lambda \neq 1),R_1,\Lambda \neq 0$%
}

In this case we have to consider roots of type I (\ref{5.20a}). We start
from the polynomial $w(\lambda )$ knowing that stable internal
space-configurations correspond to real projective coordinates $0<\lambda
<\infty $. So we have to test subpolynomials $w_{\pm }(\lambda )$ on the
existence of such roots. The high degree 
$\deg {}_\lambda [w_{\pm}(\lambda )]=2(d_1+2)(d_1+1)\geq 24$, $%
d_1\geq 2$ (because of nonvanishing curvature of the factor-spaces $M_1,M_2$%
) allows only an analysis by techniques of number theory \cite{36}, the
theory of ideals of commutative rings \cite{34} or, for general
parameter-configurations, numerical tests. In the latter case the number of
effective test-pa\-ra\-meters can be reduced by introduction of new
coordinates in parameter-space 
\mathindent=1em
\begin{eqnarray}
\label{5.25c1} 
\RR _{par}^{d_1+3} & \rightarrow & \overline{\RR }_{par}^{d_1+1} 
\!\!\!\!\!\!\!\nonumber\\ 
\overline{\RR }_{par}^{d_1+1} & = &
\left\{ (\chi ,p_i)\mid \chi =\left( 
\frac{2\Lambda d_1}{d_1+1}\right) ^{d_1+1}\frac{2P_0}{R_1^{d_1+2}},\right.
\!\!\!\!\!\!\!\nonumber\\  
 & & \left. p_i=\frac{P_i}{P_0},\ i=1,\ldots ,d_1\right\} \!\!\!\!\!\!\!
\end{eqnarray}
\mathindent=1em 
(for $P_0\neq 0,\ p_0=1$ ; in the opposite case $P_0$ can be replaced by any
nonzero $P_i$). Polynomials $w_{\pm }(\lambda )=0$ transform then to \\ $%
w_{\pm }(\lambda )=\frac 12R_1^{d_1+2}\left( (d_1+1)P_0\right) ^{d_1+1}\bar
w_{\pm }(\lambda )=0$, where 
\begin{equation}
\label{5.25c2} 
\begin{array}{l}
\bar w_{\pm }(\lambda ):= 
\widetilde{L}_3^{d_1+1}\pm (-)^{d_1+1}\chi \widetilde{b}_0^{d_1+2}=0 \\  \\ 
\widetilde{L}_3:=\bar L_3(P_0=1;P_1=p_1,\ldots ,P_{d_1}=p_{d_1}) \\  \\ 
\widetilde{b}_0:=\bar b_0(P_0=1;P_1=p_1,\ldots ,P_{d_1}=p_{d_1}). 
\end{array}
\end{equation}
Test are easy to perform with programs like {\sc mathematica} %\copyright \
or {\sc maple}. %\copyright .

As a second step we have to consider minimum conditions (\ref{5.25c003}).
Using (\ref{5.21}) we substitute 
\begin{equation}
\label{5.25d}y^{D-2}=\frac{2\bar b_0}{R_1}\ ;\qquad y^D=-\frac{(d_1+1)\bar
L_3}{\Lambda d_1} 
\end{equation}
into (\ref{5.25c}) and transform (\ref{5.25c003}) to the following
equivalent inequalities 
\begin{equation}
\label{5.25e} 
\begin{array}{l}
2(d_1+1)(d_1+2)Q_1(\lambda )+Q_2(\lambda )>\\[1ex]
>(d_1+2)(\lambda ^2+1)\bar b_0(\lambda ) \\  
\\ 
[0ex][2(d_1+2)Q_1(\lambda )-(\lambda ^2+1)\bar b_0(\lambda )]\times \\[1ex]
\times [Q_2(\lambda )-(\lambda ^2+1)\bar b_0(\lambda )]\geq \\[1ex] 
\geq (d_1+1)(\lambda ^2-1)^2\bar b_0^2(\lambda ) 
\end{array}
\end{equation}
with notations 
\mathindent=0.5em 
\begin{equation}
\label{5.25f} 
\begin{array}{l}
Q_1(\lambda ):=\sum_{i=0}^{d_1}P_i\left( \lambda ^{4+d_1+i}+\lambda
^{d_1-i}\right) \equiv \bar a_0(\lambda ) \!\!\!\!\!\!\!\!\!\!\!\!\!\!\\  
\!\!\!\!\!\!\!\\ 
Q_2(\lambda ):=\sum_{i=0}^{d_1}P_i(i+2)^2\left( \lambda ^{4+d_1+i}+\lambda
^{d_1-i}\right) . \!\!\!\!\!\!\!\!\!\!\!\!\!\!
\end{array}
\end{equation}
\mathindent=1em 

Stability-domains in parameter-space $\RR _{par}^{d_1+3}$, corresponding to
minima of the effective potential are given as intersections of domains
defined by (\ref{5.25e}) with domains which allow the existence of physical
relevant roots of $\bar w_{\pm }(\lambda )=0$. So numerical tests on minima
are easy to perform. If we additionally assume that $U_{eff}\big|
_{\min }=0$ then the class of possible stability domains narrows
considerably. Substitution of (\ref{5.25d}) into (\ref{5.25c04}) transforms
this constraint to 
\begin{equation}
\label{5.25f1}(\lambda ^2+1)\bar b_0(\lambda )=(d_1+2)Q_1(\lambda ) 
\end{equation}
and inequalities (\ref{5.25e}) to 
\begin{equation}
\label{5.25f2} 
\begin{array}{l}
d_1(\lambda ^2+1)\bar b_0(\lambda )+Q_2(\lambda )>0 \\  
\\ 
(\lambda ^2+1)[Q_2(\lambda )-(\lambda ^2+1)\bar b_0(\lambda )]\geq \\[1ex]
\geq (d_1+1)(\lambda ^2-1)^2\bar b_0(\lambda )\geq 0. 
\end{array}
\end{equation}

\subsection{Coinciding scale factors $(\lambda =1),R_1,\Lambda \neq 0$}

In this case we have to consider roots of both types. Let us start with type
I roots. Relations (\ref{5.25d}) read now 
\begin{eqnarray} 
\label{5.25ff1}
y^{D-2} & = & \frac{Q_2(1)}{R_1},\nonumber\\
y^D & = & \frac{(d_1+1)}{\Lambda d_1}\left[ Q_2(1)-2Q_1(1)\right] 
\end{eqnarray}
\mathindent=1em 
with 
\begin{eqnarray}
\label{5.25f3}
Q_1(1) & = & 2\sum_{i=0}^{d_1}P_i \\  
Q_2(1) & = & [(\lambda ^2+1)\bar b_0(\lambda )]_{\lambda
=1}=2\sum_{i=0}^{d_1}P_i(i+2)^2 \nonumber
\end{eqnarray}
so that at the extremum point the sign relation \\ $\mbox{\rm sign}(%
\sum_{i=0}^{d_1}P_i)=\mbox{\rm sign}(R_1)$ must hold and the corresponding scale factor
is simply given as 
$$
y_0=\left( \frac{2\sum_{i=0}^{d_1}P_i}{R_1}\right) ^{1/(D-2)}. 
$$
The minimum conditions (\ref{5.25e}) reduce now to 
\begin{equation}
\label{5.25g}2(d_1+2)\sum_{i=0}^{d_1}P_i>\sum_{i=0}^{d_1}P_i(i+2)^2
\end{equation}
and for $U_{eff}\big|
_{\min }=0$ even to 
\begin{equation}
\label{5.25h}(d_1+2)\sum_{i=0}^{d_1}P_i=\sum_{i=0}^{d_1}P_i(i+2)^2>0.
\end{equation}
Substitution of (\ref{5.25ff1}) into (\ref{5.25c}) shows that for $\lambda =1
$ we have 
\mathindent=-1em
\begin{eqnarray}
\label{5.25h0}
 & & \partial _{yy}^2U_{eff}\big| _{\lambda =1}  =\partial _{xx}^2U_{eff}
\big| _{\lambda =1}=\partial _{xy}^2U_{eff}\big| _{\lambda =1} \\[1ex]  
 & &  =y^{-2D+2}(d_1+1)\left[ 2(d_1+2)Q_1(1)-Q_2(1)\right] . \nonumber
\end{eqnarray}
\mathindent=1em
This implies $\det (\widetilde{A}_c)=0$ and according to (\ref{5.25c002})
the eigenvalues $w_{(c)1,2}$ of the Hessian $\widetilde{A}_c$ (\ref{5.25c001}%
) are given as 
\begin{equation}
\label{5.25h1}w_{(c)1}=2\partial _{yy}^2U_{eff}\big| _{\lambda =1},\quad
w_{(c)2}=0.
\end{equation}
The condition $w_{(c)1}>0$ is equivalent to the inequality (\ref{5.25g}). We
note that the degeneracy (\ref{5.25h0}), $\det (\widetilde{A}_c)=0,w_{(c)2}=0
$ is caused by the special and highly symmetric form of the effective
potential for the model with two identical factor-spaces. It is only in a
secondary way related to the coinciding scale factors $(x=y)$ for $\lambda
=1.$ In Appendix C we illustrate this fact with a simple counter-example,
i.e., with a potential that gives $w_{(c)1,2}>0$ for a minimum at $x=y$.

We turn now to the roots of type II given by the polynomial $I_{2+}(\lambda
=1,y)$ (\ref{5.22}). From the structure of $I_{2+}(\lambda =1)$ immediately
follows:

1. Because $I_{2+}(\lambda =1)$ contains only terms with even degree in $y$,
there exist no real roots --- and hence no extrema of the effective
potential $U_{eff}$ --- for parameter combinations with: 
\begin{equation}
\label{5.26}\mbox{\rm sign}\left( \sum_{i=0}^{d_1}P_i\right) 
=\mbox{\rm sign}(\Lambda )\neq
\mbox{\rm sign}(R_1). 
\end{equation}

2. For arbitrary parameters $\Lambda ,R_1,\bar \Delta :=\sum_{i=0}^{d_1}P_i$
roots of $I_{2+}(\lambda =1)$ can be found by analytical methods up
to dimensions $d_1\leq 2$ performing a substitution $z:=y^2$ and using
standard techniques for polynomials of degree $\deg {}_zI_{2+}(\lambda
=1)\leq 4$. Because of $R_1\neq 0\ \Leftrightarrow d_1\geq 2$ such
considerations are restricted to the case $d_1=2$.

3. There exist no general mathematical methods to obtain roots of
polynomials with degree $\deg {}_zI_{2+}(\lambda =1)>4$ and {\em arbitrary}
coefficients analytically. For special restricted classes of coefficients
techniques of number theory \cite{36} are applicable. We do not use such
techniques in the present paper. For polynomials $I_{2+}(\lambda =1)$ and
dimensions $\dim M_1=\dim M_2=d_1>2$ this implies that arbitrary parameter
sets should be analyzed numerically or parameters $\Lambda ,\ R_1,\ $ $\bar
\Delta $ should be fine tuned --- chosen ad hoc in such a way that $%
I_{2+}(\lambda =1)=0$ is fulfilled.

In the following we derive a necessary condition for the existence of a
minimum of the effective potential with fine-tuned parameters. Using the
ansatz 
\begin{equation}
\label{5.27}\frac{\Lambda d_1}{d_1+1}=\sigma _1y_0^{-2}\ ;\quad \bar \Delta
:=\sum_{i=0}^{d_1}P_i=\sigma _2y_0^{D-2} 
\end{equation}
equation (\ref{5.22}) reduces to 
\begin{equation}
\label{5.28}(\sigma _1-R_1+4\sigma _2)y_0^{D-2}=0. 
\end{equation}
Without loss of generality we choose $\sigma _2$ as free parameter, and
hence $\sigma _1=R_1-4\sigma _2$, so that from relations (\ref{5.27}) 
\begin{equation}
\label{5.28a}y_0^{D-2}=\frac{\bar \Delta }{\sigma _2},\quad y_0^D=\frac{%
d_1+1 }{\Lambda d_1}\bar \Delta (\frac{R_1}{\sigma _2}-4) 
\end{equation}
and (\ref{5.25c03}) minimum condition (\ref{5.25c02}) reads 
\mathindent=-1em
\begin{eqnarray}
\label{5.28b}
 & & \left. \partial _{yy}^2\widetilde{U}_{eff}\right| _{\min}= \\
 & & =4y_0^{-2D+2}(d_1+1)\left[ 4(d_1+2)-\frac{R_1}{\sigma _2}\right] \bar
\Delta >0 \nonumber
\end{eqnarray}
\mathindent=1em
or 
\begin{equation}
\label{5.28c}(2D-\frac{R_1}{\sigma _2})\sum_{i=0}^{d_1}P_i>0. 
\end{equation}
We see that there exists a critical value $\sigma _c=\frac{R_1}{2D}$ which
separates stability-domains with different signs of $\bar \Delta .$ From $%
y_0^{D-2}>0$ and (\ref{5.27}) follows $\mbox{\rm sign}(\bar \Delta )
=\mbox{\rm sign}(\sigma _2)$
so that 
\begin{equation}
\label{5.28d} 
\begin{array}{lll}
\bar \Delta =\sum_{i=0}^{d_1}P_i>0 & \Longleftrightarrow & \sigma _2>\sigma
_c,\quad \sigma _2>0 \\  
&  &  \\ 
\bar \Delta =\sum_{i=0}^{d_1}P_i<0 & \Longleftrightarrow & 0>\sigma
_2>\sigma _c. 
\end{array}
\end{equation}
To complete our considerations of the degenerate case $(\lambda
=1),R_1,\Lambda \neq 0$ we derive the constraint $U_{eff}\big|
_{\min }=0$. By use of (\ref{5.25c04}) and (\ref{5.28a}) this is easily done
to yield $\sigma _2=R_1/D=2\sigma _c$. So the constraint fixes the free
parameter $\sigma _2$. Remembering that according to our temporary notation $%
y:=a_2\equiv {e^{\beta ^2}\ }$ the value $y_0$ defines the scale factor of
the internal spaces at the minimum position of the effective potential, we
get now for the fine-tuning conditions (\ref{5.27}) 
\mathindent=-1em
\begin{eqnarray}
\label{5.29}
 & & \Lambda =\frac{(D-2)R_1}{Da_{(c)2}^2},\quad \bar \Delta =\frac{%
R_1a_{(c)2}^{D-2}}D,\nonumber\\ 
 & & \bar \Delta ^2=\frac{R_1^D(D-2)^{D-2}}{D^D\Lambda ^{D-2}} 
\end{eqnarray}
\mathindent=1em
--- the well-known conditions widely used in literature \cite{BZ1}. From (%
\ref{5.28d}), (\ref{5.29}) and $a_{(c)2}>0$ we see that for $\sigma
_2=R_1/D=2\sigma _c$ the stability-domain in parameter-space $\RR %
_{par}^{d_1+3}$ is narrowed to the sector 
\begin{equation}
\label{5.30}\bar \Delta =\sum_{i=0}^{d_1}P_i>0,\quad R_1>0,\quad \Lambda >0. 
\end{equation}

\subsection{Vanishing curvature-scalars $(R_1=0),\ \Lambda \neq 0$}

For vanishing curvature scalars equations (\ref{5.18}) reduce to 
\begin{eqnarray}
\label{5.35} 
I_{2+} & = & -2(D-2)\bar a_0(\lambda )-(D-4)\Lambda y^D  =  0 \quad (a)
\nonumber\\  
I_{2-} & = & -2(\lambda ^2-1)\bar b_0(\lambda )  =  0. \quad (b) 
\nonumber\\[-4ex]
 & &
\end{eqnarray}
Extrema of the effective potential are given by roots of $I_{2-}=0$ with
scale factors defined as 
\begin{equation}
\label{5.36}y^D=-\frac{2(d_1+1)\bar a_0(\lambda )}{\Lambda d_1}\equiv -\frac{%
2(d_1+1)Q_1(\lambda )}{\Lambda d_1}. 
\end{equation}
Substitution of (\ref{5.36}) into minimum-conditions (\ref{5.25c003}) yields
the following inequalities 
\mathindent=-1em
\begin{eqnarray}
\label{5.37} 
 & & Q_1(\lambda )\geq 0,\quad Q_2(\lambda )\geq 0,\quad Q_1(\lambda
)+Q_2(\lambda )>0 \!\!\!\!\!\!\!\!\!\!\!\!\!\!\!\!\nonumber\\[2ex]  
& & 8(d_1+1)(d_1+2)Q_1(\lambda )Q_2(\lambda )\geq 
 \!\!\!\!\!\!\!\!\!\!\!\!\!\!\!\!\nonumber\\[1ex]
 & & \geq (4d_1+5)^2(\lambda ^2-1)^2\bar b_0^2(\lambda ). 
\end{eqnarray}
\mathindent=1em
From (\ref{5.36}) and (\ref{5.37}) we see that for even $D$ positive $y$ are
only allowed when the bare cosmological constant $\Lambda $ is negative: $%
\Lambda <0$.

As in the case of nonvanishing curvature scalars so also roots of $I_{2-}=0$
split into two classes. For nondegenerate physical relevant configurations $%
(\lambda \neq 1)$ the corresponding $\lambda _i$ must satisfy equation 
\begin{equation}
\label{5.38}\bar b_0(\lambda )=\sum_{i=0}^{d_1}P_i(2+i)\lambda
^{d_1-i}\sum_{j=0}^{i+1}\lambda ^{2j}=0. 
\end{equation}
For $\lambda >0$ this is only possible when there exist $P_i$ with different
signs.

In the case of degenerate configurations $(\lambda =1)$ equation $I_{2-}=0$
is trivially satisfied and the scale factor at the minimum of the effective
potential given by 
\begin{equation}
\label{5.39}y_0^D=-\frac{4(d_1+1)\sum_{i=0}^{d_1}P_i}{\Lambda d_1}\ 
\end{equation}
with additional condition 
\begin{equation}
\label{5.40}
\begin{array}{l}
\sum_{i=0}^{d_1}P_i\geq 0,\quad \sum_{i=0}^{d_1}P_i(2+i)^2\geq 0,\\[1ex] 
\sum_{i=0}^{d_1}P_i\left[ (2+i)^2+1\right] >0. 
\end{array}
\end{equation}
From inequalities (\ref{5.37}) and (\ref{5.40}) immediately follows that
effective potentials with parameters $(P_0<0,\ldots ,\\ P_{d_1}<0)$ are not
stable.

\subsection{Vanishing curvature scalars and vanishing cosmological constants 
$(R_1=0,\ \Lambda =0)$}

In this case equations (\ref{5.18}) contain only the projective coordinate $%
\lambda =y/x$%
\begin{equation}
\label{5.41} 
\begin{array}{llllllll}
I_{2+} & = & -2(D-2)\bar a_0(\lambda ) & = & 0 &  &  & (a) \\  
&  &  &  &  &  &  &  \\ 
I_{2-} & = & -2(\lambda ^2-1)\bar b_0(\lambda ) & = & 0. &  &  & (b) 
\end{array}
\end{equation}
Corresponding physical configurations are possible for domains in parameter
space given by 
\begin{equation}
\label{5.42}\lambda =1,\quad \sum_{i=0}^{d_1}P_i=0 
\end{equation}
or 
\begin{equation}
\label{5.43}\lambda \neq 1,\quad R_\lambda [\bar a_0(\lambda ),\bar
b_0(\lambda )]=0. 
\end{equation}
Minima of the effective potential are localized at lines $\{\lambda _i=y/x\}$
and must be stabilized by additional terms. Otherwise we get an unstable
''run-away'' minimum of the potential.

\subsection{Generalization to $n-$scale-factor models}

The analytical methods used in the above considerations on stability
conditions of internal space configurations with two scale factors can be
extended to configurations with 3 and more scale factors by techniques of
the theory of commutative rings \cite{34}. In this case constraints, similar
to polynomial (\ref{5.11}) $w(\lambda )$, follow from resultant systems on
homogeneous polynomials. We note that in master equations (\ref{5.8}) 
we can pass from affine coordinates $\{x,\ y\}$ to projective coordinates $%
\{X, Y, Z \mid x=X/Z, y=Y/Z\}$ and transform polynomials $I_{1\pm }$
to homogeneous polynomials in $\{X,\ Y,\ Z\}$ so that these generalizations
are immediately to perform. A deeper insight in extremum conditions can be
gained by means of algebraic geometry \cite{35}. Polynomials $I_{1\pm }$
define two algebraic curves on the $\{x,\ y\}-$plane and solutions of system
(\ref{5.8}) %, (\ref{5.8a})
$I_{1\pm }(x,y)=0$ correspond to intersection-points of these curves. For $n$
scale factors extremum conditions $\{\partial _{a_i}U_{eff}=0\}_{i=1}^n$
would result in $n$ polynomials $I_i(x_1,\ldots ,x_n)=0$ defining $n$
algebraic varieties on $\RR ^n$. The sets of solutions of system $%
\{I_i(x_1,\ldots ,x_n)=0\}_{i=1}^n$, or equivalently, the intersection points of the
corresponding algebraic varieties, define the extremum points of $U_{eff}$.

\section{Conclusion}

\setcounter{equation}{0}
\renewcommand{\theequation}{\thesection\arabic{equation}}

This paper is devoted to the problem of stable compactification of internal
spaces in multidimensional cosmological models. This is one of the most
important problems in multidimensional cosmology because via stable
compactification of the internal dimensions near Planck length we can
explain unobservability of extra dimensions. With the help of dimensional
reduction we obtain an effective $D_0$-dimensional (usually $D_0=4$) theory
in the Brans-Dicke and Einstein frames. The Einstein frame is considered
here as the physical one. Stable compactification is achieved here due to the
Casimir effect which is induced by the non-trivial topology of the
space-time. The calculation of the Casimir effect in the case of more than
one scale factors is a very complicated problem. That is the reason for
proposing a Casimir-like ansatz for the energy density of the massless
scalar field fluctuations. In this ansatz all internal factors are included
on equal footing. The corresponding equation has correct Casimir dimension: $%
{[\mbox{\rm cm}]}^{-D}$ and gives correct transition to the one-scale-factor
limit. Stable configurations with respect to the internal scale factor
excitations are found in the cases of one and two internal spaces. %

\vspace{5mm}
\noindent Acknowledgement\\
\noindent 
UG acknowledges financial support from DAAD (Germany).

%%%%%%%%%%%%%%%%%%%%%%%%%%%%%%%%%%%%%%%%%%%%%%%%%%%%%%%%%
%

\section*{Appendix A: Casimir effect}

\setcounter{equation}{0} \renewcommand{\theequation}{A.\arabic{equation}} 

It
is well known that boundaries or nontrivial topology of the space-time
induce a vacuum polarization of quantized fields due to changes in the
spectrum of the vacuum fluctuations relative to a corresponding unbounded
open flat-space model. This phenomenon is known as vacuum Casimir effect and
in our case arises due to the compactness of the internal spaces $M_i\
(i=0,\ldots ,n)$. In the general case of a quantum field living at finite
temperature in a multidimensional cosmological model, the Casimir free
energy due to the presence of compactified internal factor-spaces is defined
as 
\begin{equation}
\label{A.1a}F_c=F\left[ \left\{ a_i\right\} _{i=1}^n\right] -F\left[ \left\{
a_i\rightarrow \infty \right\} _{i=1}^n\right] , 
\end{equation}
where $F\left[ \left\{ a_i\right\} _{i=1}^n\right] $ and $F\left[ \left\{
a_i\rightarrow \infty \right\} _{i=1}^n\right] $ are the free energy for a
compactified (finite-scale-factor) model and a corresponding decompactified
(flat) model, respectively. It is clear that this definition is only
meaningful, when supplemented with a regularization method yielding finite
results for the Casimir energy.

In what follows we briefly discuss some regularization techniques for the
Casimir energy of a massive scalar field in a universe with metric (\ref{2.1}%
) and give some arguments in favour of an improved version of the
Casimir-like potential, proposed in \cite{AGHK}, \cite{HKVW}.

Let us consider a massive (with the mass $M$) scalar field $\Phi $ with
arbitrary coupling to gravity in a universe with the metric (\ref{2.1}). The
Klein-Gordon-Fock equation for the conformally transformed scalar field 
\begin{equation}
\label{A.1}\Phi =\tilde \Phi {a_0}^{(1-d_0)/2} 
\end{equation}
reads \cite{Astr} 
\begin{equation}
\label{A.2} 
\begin{array}{l}
\ddot {\tilde \Phi } - \Delta \left[g^{(0)}\right]\tilde \Phi -
a_0^2\sum_{i=1}^{n} 
\frac{1}{a^2_i}\Delta \left[g^{(i)}\right]\tilde \Phi +  \\  \\ 
+ \left[M^2a^2_0 + \xi d_0(d_0-1)k_0 \right.+  \\  \\
+ \left.\xi a_0^2\sum_{i=1}^{n}\frac{%
d_i(d_i-1)k_i}{a_i^2}\right] \tilde \Phi = 0  , 
\end{array}
\end{equation}
where the overdot denotes the derivative with respect to the conformal time $%
\eta $ (we choose the time gauge: $e^\gamma =e^{\beta ^0}=a_0 $), $\xi $ is
the coupling constant, $M_i\ (i=0,\ldots ,n)$ are the spaces of constant
curvature: $R\left[ g^{(i)}\right] =k_id_i(d_i-1), \quad k_i=\pm 1,0$, and $%
\Delta \left[ g^{(i)}\right] $ are the Laplace-Beltrami operators on $M_i$: 
\begin{equation}
\label{A.3}\Delta \left[ g^{(i)}\right] =\frac 1{\sqrt{|g^{(i)}|}}\frac
\partial {\partial x^m}\left( \sqrt{|g^{(i)}|}g^{(i)mn}\frac \partial
{\partial x^n}\right) . 
\end{equation}
To get the pure Casimir effect without dynamic effects admixture we supposed
in the equation (\ref{A.2}) all scale factors to be static, what implies
that the space-time itself becomes ultrastatic.

We denote by $x^i$ the collective spatial coordinates of the i-th space and
make for the scalar field the product ansatz 
\begin{equation}
\label{A.4}\tilde \Phi =g(\eta )Y(x^0)\ldots Y(x^n)\ .
\end{equation}
Further, we assume that the scalar field on the cosmological background is
situated in the eigenstates of the Laplace-Beltrami operators 
\begin{equation}
\label{A.5}\Delta \left[ g^{(i)}\right] Y_i=-\bar n_i^2Y_i\ ,\quad
i=0,\ldots ,n\ 
\end{equation}
so that (\ref{A.2}) can be rewritten as 
$$
\ddot g+a_0^2\left[ M^2+\sum_{i=0}^n\frac{\bar n_i^2+\xi d_i(d_i-1)k_i}{a_i^2%
}\right] g=0. 
$$
Thus the physical frequency squared of the scalar field reads 
\begin{equation}
\label{A.6}\omega _{\bar n_0,\ldots ,\bar n_n}^2=M^2+\sum_{i=0}^n\frac{\bar
n_i^2+\xi d_i(d_i-1)k_i}{a_i^2}\ .
\end{equation}
The free energy of the scalar field at temperature $k_BT\equiv 1/\beta $ is
now given as \cite{KZ2}, \cite{KZ1} 
\begin{eqnarray}
\label{A.7}
F & = & \frac 1\beta \sum_J\ln {\left[ 1-\exp {\left( -\beta \hbar
\omega _{\bar n_0,\ldots ,\bar n_n}\right) }\right] }+  \nonumber\\
 & + & \frac \hbar 2\sum_J\omega _{\bar n_0,\ldots ,\bar n_n} ,
\end{eqnarray}
\mathindent=1em
where $J$ is a collective index for all quantum numbers. The eigenvalues $%
-\bar n_i^2$ depending on the topology of the spaces $M_i$ and the boundary
conditions imposed on $Y_i$ can be expressed in terms of the energy level
quantum numbers $n_i$ with degeneracy $p_i(n_i).$ For example, $\bar
n^2=n^2\quad (n=0,\pm 1,\pm 2,\ldots )\ ,\quad p=1$ for $S^1\ ;\quad \bar
n^2=n(n+1)\quad (n=0,1,2,\ldots )\ ,\quad p=2n+1$ for $S^2;$ $\bar
n^2=n^2-1\quad (n=1,2,3,\ldots )\ ,\quad p=n^2$ for $S^3$ and $\bar
n^2=n(n+3)\ ,\quad p=(n+1)(n+2)(2n+3)/6$ for $S^4$. Our further
consideration we restrict, for simplicity, to the case of the
zero-temperature (vacuum) Casimir effect. Setting $T=0$ in equation (\ref
{A.7}) we get for the vacuum free energy 
\begin{eqnarray}
\label{A.8}
F & = & \frac \hbar 2\sum_{n_0}\ldots \sum_{n_n}p_0(n_0)\ldots
p_n(n_n)\omega _{n_0,\ldots ,n_n}\equiv  \nonumber \\
 & \equiv & \frac \hbar 2\sum_J\omega _n,
\end{eqnarray}
so that the Casimir free energy according to (\ref{A.1a}) can be rewritten
as 
\begin{equation}
\label{A.9}F_c=\frac \hbar 2\sum_J\omega _n-\frac \hbar 2\int dJ\omega _n\ .
\end{equation}
The internal energy density of the fluctuations is found from the relation 
\begin{equation}
\label{A.10}\rho =\frac 1V\frac{\partial \left( \beta F_c\right) }{\partial
\beta }=\frac 1VF_c\ ,
\end{equation}
where $V=\prod_{i=0}^nV_i=\prod_{i=0}^na_i^{d_i}\mu _i$ is the space volume
of the universe and 
\begin{equation}
\label{A.11}\mu _i=\int d^{d_i}y\sqrt{|g^{(i)}|}\ ,\quad i=0,\ldots ,n\ .
\end{equation}
Pressure of fluctuations and Casimir energy-momentum tensor read,
respectively, 
\mathindent=-1em  
\begin{eqnarray}
\label{A.12}
 & & P_i=-\frac 1{\prod_{j\neq i}V_j}\frac{\partial F_c}{\partial V_i}%
=-\frac 1{Vd_i}a_i\frac{\partial F_c}{\partial a_i}\ ,\quad
i=0,\ldots ,n\ . \nonumber\!\!\!\!\!\!\!\!\\
 & &   \!\!\!\!\!\!\!\!\!
\end{eqnarray}
\mathindent=0.5em  
and 
\begin{equation}
\label{A.13}T_N^M=\mbox{\rm diag}\left( -\rho ,\underbrace{P_0,\ldots ,P_0}%
_{d_0},\ldots ,\underbrace{P_n,\ldots ,P_n}_{d_n}\right) .
\end{equation}
\mathindent=1em
Let us now turn to the consideration of regularization techniques that are
necessary for a meaningful definition of the Casimir energy (\ref{A.1a}), (%
\ref{A.9}). Most appropriate for the regularization of multiple sums like (%
\ref{A.8}) proved zeta-function technique at the one hand (for an extended
review see \cite{Eliz} and cites therein), and the multiple use of the
Abel-Plana summation formula on the other hand \cite{MT2}. For simplicity,
we perform our explicit consideration for a manifold $M$ consisting of
Ricci-flat (toroidal) factor-spaces $M=\RR \times S^1\times S^1\times \cdots
\times S^1$ . In this case the physical scalar field frequency reads 
\mathindent=-1.8em
\begin{eqnarray}
\label{A.14}
 & & \omega _J^2=M^2+4\pi ^2\left( \stackunder{external\quad space}{
\underbrace{\frac{n_{01}^2}{a_0^2}+\cdots +\frac{n_{0d_0}^2}{a_0^2}}}+%
\stackunder{internal\quad spaces}{\underbrace{\frac{n_1^2}{a_1^2}+\cdots +
\frac{n_n^2}{a_n^2}}}\right) \nonumber \!\!\!\!\!\!\!\!\\
 & & \!\!\!\!\!\!\!\!
\end{eqnarray}
\mathindent=-1em
and cummulative index and degeneracy are given as $J=\{n_{01},\ldots
,n_{0d_0},n_1,\ldots ,n_n\},$ $n_{01},\ldots ,n_n\in \ZZ $ and $%
p_J(J):=p_{01}(n_{01})\ldots p_n(n_n)=1.$ Because of $p_J(J)=1$ the free
energy (\ref{A.8}) can be expressed in terms of analytically continued
inhomogeneous Epstein zeta-functions 
\begin{eqnarray}
\label{A.15}
 & & Z_N^M(s;b_1,\ldots ,b_N):= \nonumber \\
 & & :=\sum_{n_1,\cdots ,n_N\in \ZZ }\left(
M^2+b_1n_1^2+\cdots +b_Nn_N^2\right) ^{-s}
\end{eqnarray}
and 
\begin{eqnarray}
\label{A.16}
 & & E_N^M(s;b_1,\ldots ,b_N):= \nonumber \\ 
 & & :=\sum_{n_1,\cdots ,n_N=1}^\infty \left(
M^2+b_1n_1^2+\cdots +b_Nn_N^2\right) ^{-s}
\end{eqnarray}
\mathindent=1em
directly \cite{Eliz} 
\begin{eqnarray}
\label{A.17}
F & = & \frac \hbar 2\sum_{n_{01},\cdots ,n_n=-\infty }^\infty \omega
_{n_{01},\ldots ,n_n}= \nonumber\\
 & = & \frac \hbar 2Z_N^M\left( s\rightarrow -\frac
12;a_0^{-2},\ldots ,a_0^{-2},a_1^{-2},\ldots ,a_n^{-2}\right)  
\!\!\!\!\!\!\!\!\nonumber \\  
 & = & \frac \hbar 2\sum_{j=0}^{N-1}2^{N-j}\sum_{\left\{ k_1,\ldots ,k_j\right\}
}E_{N-j}^M\left( s\rightarrow -\frac 12;\right. 
\!\!\!\!\!\!\!\!\nonumber \\
 & & \left.a_0^{-2},\ldots ,\widehat{%
a_{k_1}^{-2}},\ldots ,\widehat{a_{k_j}^{-2}},\ldots ,a_n^{-2}\right) ,
\end{eqnarray}
where $\left\{ k_1,\ldots ,k_j\right\} $ denotes a combination of $j$
indices --- whose corresponding terms are absent --- out of $N\equiv d_0+n$
possible ones and we have included the prefactor $4\pi ^2$ into the scale
factors $a_i/2\pi \rightarrow a_i$ for convenience. (The absent terms $
\widehat{a_{k_j}^{-2}}$ result from the zeros of the summation indices $%
n_{k_j}=0$ when we change the summation sets from $n_{01},\ldots ,n_n\in \ZZ 
$ to $n_{01},\ldots ,n_n\in \left\{ 1,2,3,\ldots \right\} $ passing from $%
Z_N^M$ to $E_N^M.$) Multiple iterative use of the exact formula \cite{Eli1} 
\mathindent=-1.8em
\begin{eqnarray}
\label{A.18}
 & & E_N^M(s;b_1,\ldots ,b_N)=-\frac 12E_{N-1}^M(s;b_2,\ldots ,b_N)+ 
 \nonumber\\
 & & \frac 12
\sqrt{\frac \pi {b_1}}\frac{\Gamma (s-\frac 12)}{\Gamma (s)}%
E_{N-1}^M(s-\frac 12;b_2,\ldots ,b_N)+ \nonumber\\   
 & & +
\frac{\pi ^s}{\Gamma (s)}b_1^{-\frac s2}\sum_{k=0}^\infty \frac{b_1^{k/2}}{%
k!(16\pi )^k}\prod_{j=1}^k\left[ \left( 2s-1\right) ^2-\left( 2j-1\right)
^2\right] \times \nonumber\\
 & & \times \sum_{n_1,\ldots ,n_N=1}^\infty n_1^{s-k-1}\times  \nonumber\\  
 & & \times 
 \left( M^2+b_2n_2^2+\cdots +b_Nn_N^2\right) ^{-(s+k)/2} \times 
\nonumber\\
 & & \times \exp \left[ -
\frac{2\pi }{\sqrt{b_1}}\left( M^2+b_2n_2^2+\cdots +b_Nn_N^2\right)
^{1/2}\right] 
\end{eqnarray}
\mathindent=-1em
with start condition for the iteration 
\begin{eqnarray}
\label{A.19}
 & & E_1^M(s;1)  =  -\frac{M^{-2s}}2+\frac{\sqrt{\pi }}2\frac{\Gamma
(s-\frac 12)}{\Gamma (s)}M^{-2s+1} + \nonumber\\
 & & +  \frac{2\pi ^sM^{-s+1/2}}{\Gamma (s)}%
\sum_{p=1}^\infty p^{s-1/2}K_{s-1/2}(2\pi pM)
\end{eqnarray}
would allow to express the Casimir energy in terms of exponential functions
and modified Bessel functions $K_\nu $ of the second kind. With these
formulae at hand an analysis of the asymptotic behavior of the Casimir
energy can be performed. For the case of two scale factors with $a_2\leq a_1$
the corresponding Epstein zeta functions after an additional regularization
have been estimated in Ref. \cite{Eli2} to yield 
\begin{eqnarray}
\label{A.20}
 & & 
\sum_{n_1,n_2=1}^\infty \sqrt{\left( \frac{n_1}{a_1}\right) ^2+\left( \frac{%
n_2}{a_2}\right) ^2} = \\
& & =\frac 1{24}\left( \frac 1{a_1}+\frac 1{a_2}\right) -
\frac{\varsigma (3)}{8\pi ^2}\left( \frac{a_2}{a_1^2}+\frac{a_1}{a_2^2}%
\right) - \nonumber\\    
 & & -\frac{\pi ^{3/2}}{2\sqrt{a_1a_2}}\exp 
 \left( -2\pi \frac{a_1}{a_2}\right)
\left[ 1+O(10^{-3})\right] \nonumber
\end{eqnarray}
and 
\begin{eqnarray}
\label{A.21}
 & & \sum_{n_1,n_2=1}^\infty \sqrt{\left( \frac{n_1}{a_1}\right) ^2+\left( \frac{%
n_2}{a_2}\right) ^2+M^2}  = \\
 & & =\frac M4-\frac \pi 6a_1a_2M^3+\left( \frac
1{4\pi }
\sqrt{\frac M{a_2}}-\frac{Ma_1}{4\pi a_2}\right) \times  \nonumber\\  
 & & \times \exp \left( -2\pi Ma_2\right) \left[ 1+O(10^{-3})\right] .
 \nonumber
\end{eqnarray}
\mathindent=1em 
To our knowledge, for three and more scale factors similar estimates are
performed up to now with less precission (see e.g. \cite{MT2}). The
calculations become more complicated, when one tries to develop analytical
approaches for compactified non-Ricci-flat internal factor-spaces with more
then one scale factor. In this case the nontrivial degeneracy factor $%
p_J(J)\neq 1$ in the sums of type (\ref{A.8}) prevents a direct use of
Epstein zeta function methods. Examples for an analytical circumvention of
this problem by use of resummation and Mellin transformation are given in 
\cite{Eliz} for the case of different one-scale-factor-spaces.
Generalizations to multi-scale-factor models will need additional efforts in
future.

For completeness we make now some brief remarks on the regularization method
based on the Abel-Plana summation formula \cite{Erd}, \cite{Evg}, \cite{Olv} 
\begin{eqnarray}
\label{A.22}\sum_{n=0}^\infty f(n) & = & \frac{f(0)}2+\int\limits_0^\infty
f(n)dn+ \nonumber\\
 & + & i\int\limits_0^\infty \frac{f(i\nu )-f(-i\nu )}{\exp {(2\pi \nu )}-1}%
d\nu \ 
\end{eqnarray}
for a function $f(n)$ satisfying some conditions in the complex plane $\CC $
. This method was developed mainly in Refs. \cite{MT1}, \cite{MT2} and
consists according to (\ref{A.9}) in a subtraction of divergent terms
containing $\int\limits_0^\infty f(n)dn.$ It yields the Casimir energy as
finite expression. We applied this method, e.g., to get low-temperature
corrections to the vacuum Casimir effect caused by scalar fields living, for
example, in a closed Friedmann universe \cite{KZ2} and on a manifold with
three-dimensional torus topology $M=\RR \times S^1\times S^1\times S^1$ \cite
{KZ1} (for a consideration of the high temperature regime with the help of
zeta function techniques see \cite{kirsten}). Direct application of (\ref
{A.22}) to the triple sum (\ref{A.9}) gives, e.g., for a massless scalar
field \cite{KZ1} 
\mathindent=-1em
\begin{eqnarray}
\label{A.23}
& & F_c  =  \frac{4\pi \hbar}{3}\left\{-\int \limits_{0}^{\infty}d\nu \,
I\!\left(\frac{\nu}{a};0,0\right) + \right. \nonumber\\
& & +  \int \limits_{0}^{\infty}dl\left[\int
\limits_{\frac{b}{a}l}^{\infty}d\nu \,
I\!\left(\frac{\nu}{b};\frac{l}{a},0\right) + \int
\limits_{\frac{c}{a}l}^{\infty}d\nu \,
I\!\left(\frac{\nu}{c};\frac{l}{a},0\right) \right] + \nonumber\\
& & +  2\sum_{l=0}^{\infty}\left[\int
\limits_{\frac{b}{a}l}^{\infty}d\nu \,
I\!\left(\frac{\nu}{b};\frac{l}{a},0\right) + \int
\limits_{\frac{c}{a}l}^{\infty}d\nu \,
I\!\left(\frac{\nu}{c};\frac{l}{a},0\right) \right] - \nonumber\\
& & -  4\int \limits_{0}^{\infty}\int \limits_{0}^{\infty}dn\,dm\int
\limits_{a\sqrt{\frac{m^2}{b^2}+\frac{n^2}{c^2}}}^{\infty}d\nu \,
I\!\left(\frac{\nu}{a};\frac{m}{b},\frac{n}{c}\right) - \nonumber\\
& & -  2\int \limits_{0}^{\infty}dn\left(\sum_{l=0}^{\infty}\int
\limits_{b\sqrt{\frac{l^2}{a^2}+\frac{n^2}{c^2}}}^{\infty}d\nu \,
I\!\left(\frac{\nu}{b};\frac{l}{a},\frac{n}{c}\right)\right. +
\nonumber\\
& & +  \left.\sum_{m=0}^{\infty}\int
\limits_{a\sqrt{\frac{m^2}{b^2}+\frac{n^2}{c^2}}}^{\infty}d\nu \,
I\!\left(\frac{\nu}{a};\frac{m}{b},\frac{n}{c}\right)\right) -
\nonumber\\
& & -  4\left.\sum_{l,m=0}^{\infty}\int
\limits_{c\sqrt{\frac{l^2}{a^2}+\frac{m^2}{b^2}}}^{\infty}d\nu \,
I\!\left(\frac{\nu}{c};\frac{l}{a},\frac{m}{b}\right) + C.P.\right\}\ ,
\end{eqnarray}
\mathindent=0.5em 
where 
\begin{equation}
\label{A.24}I\negthinspace \left( \frac \nu a;\frac mb,\frac nc\right) = 
\sqrt{\frac{\nu ^2}{a^2}-\frac{m^2}{b^2}-\frac{n^2}{c^2}}{\left( e^{2\pi \nu
}-1\right) }^{-1}, 
\end{equation}
\mathindent=1em
$a,b,c$ are the scale factors corresponding to the wrapped factor-spaces of
the three-torus, and $C.P.$ means terms obtained by cyclic permutations $%
\{a,l;b,m;c,n\}$, for example, $b\frac nc\rightarrow c\frac la\rightarrow
a\frac mb$. Similar to the derivation of (\ref{A.20}) the symmetric entering
of the factor-space scale factors $a,b,c$ into (\ref{A.23}) will be
recovered explicitly only at a final stage of an asymptotic expansion. For
some considerations on a possible use of an Abel-Plana summation formula
generalized to $\CC ^n$ we refer to Appendix B.

From the fact that the scale factors for factor-spaces of the same
topological type (the same dimension and curvature) up to higher order
corrections enter into the physical frequency according to (\ref{A.6}) in
the same way we conclude that there must exist a corresponding interchange
symmetry in the expression for the Casimir energy. Asymptotical expansion (%
\ref{A.20}) for the toroidal two-scale-factor model shows this symmetry
explicitly. Furthermore, from the one-scale-factor model that effectively
emerges in the case of coinciding scale factors $a_1=a_2=\ldots =a_n=a$ we
know that the Casimir energy density in this case has the form 
\begin{equation}
\label{A.25}\rho =Ce^{-D\beta }=\frac C{a^D}\ , 
\end{equation}
where $D=1+d_0+\sum_{i=1}^nd_i$ is the dimension of the total space-time and 
$C$ is a dimensionless constant which depends strongly on the topology of
the model. It is clear that the Casimir energy density has the same
structure for models with only one scale factor \cite{Ford}, \cite{MMS}, 
\cite{MT1}, \cite{MT2}, \cite{KZ2} (or all scale factors identically equal
each other \cite{DeW} : $a_0\equiv a_1\equiv \dots \equiv a_n\equiv a$ ).
Explicit calculations, formula (\ref{A.20}) from \cite{Eli2} and asymptotic
expansion 
\begin{equation}
\label{A.26}\rho =-\frac{\pi ^2}{90a^4}-\frac{\zeta (3)}{2\pi ab^3}-\frac
\pi {6abc^2}\ 
\end{equation}
derived in papers \cite{MT1}, \cite{MT2} for the case $c\geq b\geq a$ show
that the Casimir energy density as function of the scale factors has an
effective degree $\deg (\rho )=-D$ and in a serial expansion of $\rho $
terms containing $a_i^{-\widetilde{D}},$ $\widetilde{D}>D$ are compensated
by factors containing other scale factors. A sufficiently simple function
with this feature and which under decompactification of some of the factor-%
spaces yields the correct new degree can be constructed as 
\mathindent=0.5em
\begin{eqnarray}
\label{A.27}
\rho  & = & e^{-\sum_{i=0}^nd_i\beta ^i}\sum_{k_0,\ldots
,k_n=0}^n\epsilon _{k_0|k_1\ldots k_n|}\sum_{\xi _0=0}^{d_{k_1}}\sum_{\xi
_1=0}^{d_{k_2}}\ldots \nonumber\\
 & \ldots & \sum_{\xi _{n-1}=0}^{d_{k_n}}A_{\xi _0\ldots \xi
_{n-1}}^{(k_0)}\frac{{\left( e^{\beta ^{k_1}}\right) }^{\xi _0}\ldots {%
\left( e^{\beta ^{k_n}}\right) }^{\xi _{n-1}}}{{\left( e^{\beta
^{k_0}}\right) }^{\xi _0+\xi _1+\cdots +\xi _{n-1}+1}}  , \nonumber\\ 
 & & 
\end{eqnarray}
where $A_{\xi _0\ldots \xi _{n-1}}^{(k_0)}$ are dimensionless constants
which depend on the topology of the model and $\epsilon _{ik\ldots m}={(-1)}%
^P\varepsilon _{ik\ldots m}$. Here $\varepsilon _{ik\ldots m}$ 
is the totally antisymmetric symbol
($\varepsilon _{01\ldots m}=+1$), $P$ is the number of the 
permutations of the $%
01\ldots n$ resulting in $ik\ldots m$. $|k_1k_2\ldots k_n|$ means summation
is taken over $k_1<k_2<\cdots <k_n$. Formally, the Eq. (\ref{A.27}) follows
from expression (\ref{A.20}) if we generalize it to the case of $M=\RR %
\times T^{d_0}\times \cdots \times T^{d_n},$ where $T^{d_i}$ are 
$d_i$-tori,
and omit the exponential terms.

Let $M_0$ denote the external space and its scale factor $a_0\rightarrow
\infty $. Then equation (\ref{A.27}) yields 
\begin{eqnarray}
\label{A.28}
\rho & = & e^{-\sum_{i=1}^nd_i\beta ^i}\sum_{k_1,\ldots
,k_n=1}^n\epsilon _{k_1|k_2\ldots k_n|}\sum_{\xi _1=0}^{d_{k_2}}\ldots
\nonumber\\ 
 & \ldots & \sum_{\xi _{n-1}=0}^{d_{k_n}}A_{\xi _1\ldots %
\xi _{n-1}}^{(k_1)}\frac{{%
\left( e^{\beta ^{k_2}}\right) }^{\xi _1}\ldots {\left( e^{\beta
^{k_n}}\right) }^{\xi _{n-1}}}{{\left( e^{\beta ^{k_1}}\right) }^{D_0+\xi
_1+\cdots +\xi _{n-1}}}\ , \nonumber\\ 
 & &
\end{eqnarray}
where $A_{\xi _1\ldots \xi _{n-1}}^{(k_1)}\equiv A_{d_0\xi _1\ldots \xi
_{n-1}}^{(k_1)}$ and $D_0=d_0+1$. For example, in the case of the two
internal spaces and $a_0\gg a_1,a_2$ Eq. (\ref{A.27}) reads 
\mathindent=0.2em
\begin{equation}
\label{A.29}\rho =\frac 1{a_1^{d_1}a_2^{d_2}}\left[ \sum_{\xi =0}^{d_2}A_\xi
^{(1)}\frac{a_2^\xi }{a_1^{D_0+\xi }}+\sum_{\xi =0}^{d_1}A_\xi ^{(2)}\frac{%
a_1^\xi }{a_2^{D_0+\xi }}\right] . 
\end{equation}
\mathindent=1em
It is clear, that for identical spaces $M_1\equiv M_2$ we have: $%
d_1=d_2,R\left[ g^{(1)}\right] =R\left[ g^{(2)}\right] $ and $A_\xi
^{(1)}=A_\xi ^{(2)}$.

Equations (\ref{A.27}), (\ref{A.28}) and (\ref{A.29}) have correct Casimir
dimension: $\mbox{\rm cm}^{-D}$ and give correct transition to the case of
one scale factor (\ref{A.25}). This can be easily seen, for example, from
Eq. (\ref{A.29}) when one of the scale factors goes to infinity. In one's
turn equation (\ref{A.29}) may be obtained from equation (\ref{A.28}) if $%
n-2 $ internal scale factors go to infinity. Thus, in contrast to the
expression (\ref{1.3}) for the Casimir energy density proposed in the papers 
\cite{AGHK}, \cite{HKVW} in our case the energy density does not equal to
zero if at least one of the scale factors is finite.

In the case of a large external space: $a_0\gg a_1,\ldots ,a_n$ from the
equations (\ref{A.10}), (\ref{A.12}) and (\ref{A.28}) we can easily derive
the equation of state in the external space: 
\begin{equation}
\label{A.30}P_0=-\rho \ , 
\end{equation}
because here $F_c=V\rho =\left( \prod_{i=0}^n\mu _i\right) e^{\gamma _0}\rho
\left( \beta ^1,\ldots ,\beta ^n\right) $ and $\gamma
_0=\sum_{i=0}^nd_i\beta ^i$, and 
\begin{equation}
\label{A.31}\sum_{i=1}^nd_iP_i=D_0\rho . 
\end{equation}

\section*{Appendix B: Abel-Plana summation formula in $\CC ^n$}

\setcounter{equation}{0} \renewcommand{\theequation}{B.\arabic{equation}}

Here we give some arguments for a generalization of the Abel-Plana summation
formula (\ref{A.22}) to the case of several variables. For this purpose we
first recall that (\ref{A.22}) is the result of a contour integration in $%
\CC .$ Given a contour $\gamma =\partial A\subset \CC $ as boundary of a
rectangle $A=\left\{ z\in \CC \mid \left| Im(z)\right| \leq R;a\leq
Re(z)\leq b\right\} $ one considers the Cauchy contour integral 
\begin{equation}
\label{B.1}\oint_\gamma f(z)\mbox{\rm ctg}(\pi z)dz. 
\end{equation}
If necessary the contour $\gamma $ should be deformed appropriately to pass
around possible branch cuts of $f(z).$ Adding $\oint_{\partial
A_{+}}f(z)dz-\oint_{\partial A_{-}}f(z)dz$, where $A_{\pm }=\left\{ z\in \CC %
\left| 
\begin{array}{l}
0\leq Im(z)\leq R \\ 
0\geq Im(z)\geq R 
\end{array}
\right. ;a\leq Re(z)\leq b\right\} $ are the halfs of the rectangle $%
A=A_{+}\cup A_{-}$ located at the upper and the lower complex half-plane, we
arrive at 
\begin{eqnarray}
\label{B.2}
\sum_{s\in L_A}f(s) & = & \int_a^bf(z)dz+\int_{\gamma _{+}}\frac{f(z)dz}{%
1-e^{-2\pi iz}}+ \nonumber \\
 & + & \int_{\gamma _{-}}\frac{f(z)dz}{e^{2\pi iz}-1}. 
\end{eqnarray}
Here the integration contours are defined as $\gamma _{\pm }=\left\{ z\in
\gamma \left| 
\begin{array}{l}
0\leq Im(z) \\ 
0\geq Im(z) 
\end{array}
\right. \right\} $ and $L_A=(a,b)\cap \ZZ $. To get Eq. (\ref{A.22}) the
conditions $\left| f(iR+t)\right| \leq \varphi (t)e^{\alpha \left| R\right|
},\alpha <2\pi $; $\int_a^b\varphi (t)dt\leq M<\infty $ and $\varphi (t)%
\stackunder{t\rightarrow \infty }{\longrightarrow }0$ should be satisfied.
This means that for an application of the Abel-Plana summation formula to
the calculation of the Casimir energy the physical frequency should be
multiplied by an appropriate cutoff function. At the end of the calculations
this cutoff can be removed.

Relation (\ref{B.2}) is well suited for a generalization to a multi-contour
integral in $\CC ^n.$ Following \cite{fu,ci} we start with a function $f(z)$
polyholomorphic in the domain $S:=A_1\times A_2\times \cdots \times
A_n\subset \CC ^n$ built from the rectangles $A_i\subset \CC _i^1.$ Then for 
$f(z)$ holds the integral representation 
\mathindent=-1em  
\begin{eqnarray}
\label{B.3}
 & & f(z)= \\
 & & \frac 1{(2\pi i)^n}\oint_{\partial S}\frac{f(t)}{%
(t_1-z_1)\cdot \cdots \cdot (t_n-z_n)}dt_1\wedge \cdots \wedge dt_n, 
\nonumber
\end{eqnarray}
where $\partial S$ is defined as $\partial S:=\partial A_1\times \partial
A_2\times \cdots \times \partial A_n$. Making use of $\sum_{k=-\infty
}^\infty \frac 1{a-k}=\pi \mbox{\rm ctg}(\pi a)$ we can 
extend this representation to 
\mathindent=-1.8em
\begin{eqnarray}
\label{B.4}
 & & \sum_{\left\{ k_1,\ldots ,k_n\right\} \in L_S}f(k_1,\ldots
,k_n) = \\
 & & \frac 1{(2\pi i)^n}\oint_{\partial S}f(z_1,\ldots ,z_n) 
 \prod_{j=1}^n\left[ \pi \mbox{\rm ctg}(\pi z_j)\right] dz_1\wedge 
\cdots \wedge dz_n \!\!\!\!\!\!\!\!\!\!\!\!\!\!\!\!\nonumber
\end{eqnarray}
\mathindent=1em
with $L_S=S\cap \ZZ ^n$ --- the $n-$dimensional lattice segment contained in
the domain $S$. Proceeding as in the case of one complex variable, a
relation generalizing (\ref{B.2}) can be obtained and convergence criteria
can be established. The main task in considerations for functions $%
f(z_1,\ldots ,z_n)$ with branch cuts, as in the case of the Casimir energy,
consists in nontrivial contour deformations and in keeping control over
signs and sign changes of the integrals when moving along the oriented
contours.

As in the calculations of the Casimir energy in Refs. \cite{MT1}, \cite{MT2}
step by step calculations with the help of (\ref{B.4}) lead to expressions
containing terms with some functions $g$ of the form $g(\frac{a_i}{a_k}z_l)$
. These functions lead to an aparent asymmetry between the scale factors $%
a_i $ at intermedium stages of the calculation, as explicitly shown, e.g.,
in (\ref{A.23}) and \cite{Eliz} also. Nevertheless, final asymptotic
expansions can be transformed to expressions symmetric in the scale factors.

\section*{Appendix C: 
Potential with minimum of type $w_{(c)1,2}>0$ for
coinciding scale factors}

\setcounter{equation}{0} \renewcommand{\theequation}{C.\arabic{equation}}

In this Appendix we show with the help of a simple example that a minimum of
an effective potential $U_{eff}(x,y)$ at a point with coinciding scale
factors $(x=y)$ implies not necessarily a degenerate Hessian $\widetilde{A}_c
$ (\ref{5.25c001}) $\det (\widetilde{A}_c)=0,w_{(c)1}>0,w_{(c)2}=0$ . As
potential we choose 
\begin{equation}
\label{c.1}U_{eff}(x,y)=f(x)g(y)+h(xy)
\end{equation}
with 
\begin{equation}
\label{c.2}
\begin{array}{l}
f(x)=a_{-}x^{-1}+a_{+}x \\  
\\ 
g(y)=b_{-}y^{-1}+b_{+}y \\  
\\ 
h(xy)=c_{-}(xy)^{-1}+c_{+}xy.
\end{array}
\end{equation}
This ansatz is not related to any of the physical potentials considered as
approximation to the exact Casimir potential. But we will see that for some
coefficient sets $\left\{ a_{\pm },b_{\pm },c_{\pm }\right\} $ it provides a
minimum with $w_{(c)1,2}>0$ for coinciding scale factors $x=y>0$. Proceeding
as in subsection \ref{mark1} we start from the extremum condition $\partial
_x$$U_{eff}=0,\quad \partial _yU_{eff}=0$ and consider the equation system $%
x\partial _x$$U_{eff}\pm y\partial _yU_{eff}=0.$ For $x=y$ this yields 
\begin{equation}
\label{c.3}
\begin{array}{l}
-(a_{-}b_{-}+c_{-})y^{-3}+(a_{+}b_{+}+c_{+})y^{-1}=0 \\  
\\ 
(-a_{-}b_{+}+a_{+}b_{-})y^{-1}=0
\end{array}
\end{equation}
so that for $y>0$ we have 
\begin{equation}
\label{c.4}y^4=\frac{a_{-}b_{-}+c_{-}}{a_{+}b_{+}+c_{+}},\qquad
a_{+}b_{-}=a_{-}b_{+}
\end{equation}
and the second derivatives of the potential read 
\mathindent=-1em
\begin{eqnarray}
\label{c.5}
 & & \partial _{xx}U_{eff}=  \partial _{yy}U_{eff}  
=2(a_{+}b_{+}+c_{+})+2a_{-}b_{+}y^{-2} 
\!\!\!\!\!\!\!\!\!\!\!\!\!\!\!\!\nonumber\\  
& & \partial _{xy}U_{eff} =2(a_{+}b_{+}+c_{+})-2a_{-}b_{+}y^{-2}.
\end{eqnarray}
\mathindent=1em
Hence the eigenvalues of the Hessian $\widetilde{A}_c$ are given as $%
w_{(c)1,2}=\partial _{yy}U_{eff}\pm \partial _{xy}U_{eff}$ or $%
w_{(c)1}=4(a_{+}b_{+}+c_{+})$ and $w_{(c)1,2}=4a_{-}b_{+}y^{-2}$. Choosing
appropriate nonvanishing coefficients $a_{\pm },b_{\pm },c_{\pm }$ the
minimum conditions $w_{(c)1,2}>0$ are easy to satisfy so that in the general
case the potential $U_{eff}$ can have a nondegenerate minimum at a point
with coinciding scale factors $x=y$.

%%%%%%%%%%%%%%%%%%%%%%%%%%%%%%%%%%%%%%%%%%%%%%%%%%%%%%%%%%%%%%%%%
%

%
%%%%%%%%%%%%%%%%%%%%%%%%%%%%%%%%%%%%%%%%%%%%%%%%%%%%%%%%%%%%%%%%%

\end{document}